\documentclass[aps,prc,twocolumn,groupedaddress,nofootinbib]{revtex4-2}
\usepackage{bm}
\usepackage[dvipdfmx]{graphicx}
\usepackage{ascmac}
\usepackage{amsmath,amssymb}
\usepackage{braket}
\usepackage{url}
\usepackage{mathrsfs}
\usepackage{cancel}
\allowdisplaybreaks[1]

\usepackage[normalem]{ulem}  % \sout{old text} for strikeout
\usepackage{color} % For blue in-text comments and additions

\renewcommand\sout{\bgroup \color{red} \ULdepth=-.5ex \ULset}

\begin{document}

% Use the \preprint command to place your local institutional report
% number in the upper righthand corner of the title page in preprint mode.
% Multiple \preprint commands are allowed.
% Use the 'preprintnumbers' class option to override journal defaults
% to display numbers if necessary
%\preprint{}

%Title of paper
\title{Analysis of $\Xi(1620)$ resonance and $\bar{K}\Lambda$ scattering length with chiral unitary approach}

% repeat the \author .. \affiliation  etc. as needed
% \email, \thanks, \homepage, \altaffiliation all apply to the current
% author. Explanatory text should go in the []'s, actual e-mail
% address or url should go in the {}'s for \email and \homepage.
% Please use the appropriate macro foreach each type of information

% \affiliation command applies to all authors since the last
% \affiliation command. The \affiliation command should follow the
% other information
% \affiliation can be followed by \email, \homepage, \thanks as well.
\author{Takuma Nishibuchi}
\email[]{nishibuchi-takuma@ed.tmu.ac.jp}
%\homepage[]{Your web page}
%\thanks{}
%\altaffiliation{}
\affiliation{Department of Physics, Tokyo Metropolitan University, Hachioji 192-0397, Japan}

\author{Tetsuo Hyodo}
\email[]{hyodo@tmu.ac.jp}
%\homepage[]{Your web page}
%\thanks{}
%\altaffiliation{}
\affiliation{Department of Physics, Tokyo Metropolitan University, Hachioji 192-0397, Japan}

%Collaboration name if desired (requires use of superscriptaddress
%option in \documentclass). \noaffiliation is required (may also be
%used with the \author command).
%\collaboration can be followed by \email, \homepage, \thanks as well.
%\collaboration{}
%\noaffiliation

\date{\today}

\begin{abstract}
We study the $\Xi(1620)$ resonance near the $\bar{K}\Lambda$ threshold in the light of the recent experimental constraints. The Belle collaboration have found a resonance peak of $\Xi(1620)$ slightly below the $\bar{K}^{0}\Lambda$ threshold in the $\pi^{+}\Xi^{-}$ invariant mass spectrum, and the ALICE collaboration have determined the $K^{-}\Lambda$ scattering length from the measurement of the momentum correlation functions in the heavy ion collisions. Using the effective range expansion, we classify the nature of the pole of the near-threshold eigenstate in terms of the scattering length, in the presence of the decay channel. It is shown that the quasibound state below the threshold can be described by only the scattering length, while the description of the resonance above the threshold requires the contribution from the effective range. Based on the chiral unitary approach, we construct a theoretical model which generates the pole of $\Xi(1620)$ below the $\bar{K}\Lambda$ threshold with relatively narrow width, as reported by the Belle collaboration. It is quantitatively demonstrated that the spectrum of the $\Xi(1620)$ quasibound state is distorted by the effect of the nearby $\bar{K}\Lambda$ threshold. We then construct another model which reproduces the $K^{-}\Lambda$ scattering length by the ALICE collaboration. In this case, the eigenstate pole does not appear in the physically relevant Riemann sheets, and the spectrum shows a cusp structure at the $\bar{K}\Lambda$ threshold. We finally examine the compatibility of the value of the $\bar{K}\Lambda$ scattering length and the subthreshold pole of $\Xi(1620)$ including the experimental uncertainties.
\end{abstract}

% insert suggested keywords - APS authors don't need to do this
%\keywords{}

%\maketitle must follow title, authors, abstract, and keywords
\maketitle

% body of paper here - Use proper section commands
% References should be done using the \cite, \ref, and \label commands
\section{Introduction}

% introduction of Xi
The $\Xi$ baryon (strangeness $S=-2$, isospin $I=1/2$) has two charged components, $\Xi^0$ and $\Xi^-$, constructed by $uss$ and $dss$, respectively. While almost thirty excited states are confirmed in the nucleon sector experimentally, only about ten excited $\Xi$ baryons have been established so far~\cite{ParticleDataGroup:2022pth,Hyodo:2020czb}. Furthermore, the nature of the most of excited $\Xi$ states is not well understood. This is because the study of the excited $\Xi$ states requires the double strangeness production, and therefore experimental data were not accumulated very much. In theoretical analysis, various model calculations have been performed, but the nature of the excited $\Xi$ states have not been well clarified, due to the lack of experimental data.

% recent experiments
Recently, there have been great progress in obtaining the detailed data of the low-lying $\Xi$ excited states experimentally. In 2019, the Belle collaboration observed peaks of $\Xi(1620)$ and $\Xi(1690)$ in the invariant mass distribution of $\pi^+\Xi^-$ in the $\Xi_{c}\to \Xi^{-}\pi^{+}\pi^{+}$ decay~\cite{Belle:2018lws}. In 2020, the $\Xi(1690)$ resonance was observed in the $\Lambda K^-$ invariant mass distribution in the $\Xi_b\rightarrow J/ \psi\Lambda K^-$ decay by the LHCb collaboration~\cite{LHCb:2020jpq}. In 2021, through the measurement of the two-body momentum correlation function in the Pb-Pb collisions, the ALICE collaboration reported the scattering length of $K^-\Lambda$ whose threshold energy is  close to the mass of $\Xi(1620)$~\cite{ALICE:2020wvi}. It is expected that these detailed data will serve to elucidate the nature of $\Xi(1620)$ and $\Xi(1690)$.

% theoretical studies
Theoretical studies have been performed mainly before these recent experimental developments. For instance, in the constituent quark model~\cite{Isgur:1978xj}, the masses of $\Xi(1620)$ and $\Xi(1690)$ are predicted to be much higher than the those in Belle and LHCb results. A study by the lattice QCD calculation with $m_{\pi}\geq 225$ MeV have obtained the reasonable effective masses of the ground state and $\Xi(1535)$, but the signals of other excited states are not conclusive~\cite{Engel:2013ig}. In contrast to these static frameworks, the chiral unitary approach~\cite{Kaiser:1995eg,Oset:1997it,Oller:2000fj,Hyodo:2011ur} dynamically generates the excited $\Xi$ states as resonances in the meson-baryon scattering. In Ref.~\cite{Ramos:2002xh}, the mass of $\Xi(1620)$ is obtained at around 1600 MeV, but with a broad width. It is shown that $\Xi(1620)$ and $\Xi(1690)$ can be generated simultaneously in the work of Ref.~\cite{Garcia-Recio:2003ejq}. Additionally, this work has been expanded to an SU(6) model in Ref.~\cite{Gamermann:2011mq}. By focusing on $\Xi(1690)$, Ref.~\cite{Sekihara:2015qqa} generates a pole near the $\bar{K}\Sigma$ threshold as a hadronic molecule state. It is shown that the model can reproduce the invariant mass distribution of $\bar{K}\Sigma$ and $\bar{K}\Lambda$ in the $\Lambda_c$ decay~\cite{Belle:2001hyr}. A recent work~\cite{Feijoo:2023wua} after the Belle measurement indicates that $\Xi(1620)$ and $\Xi(1690)$ can be reproduced simultaneously by taking into account the next to leading order chiral interaction, but again with a broad decay width of about 150 MeV for $\Xi(1620)$. 

\begin{figure}[tbp]
    \centering
    \includegraphics[width=8cm]{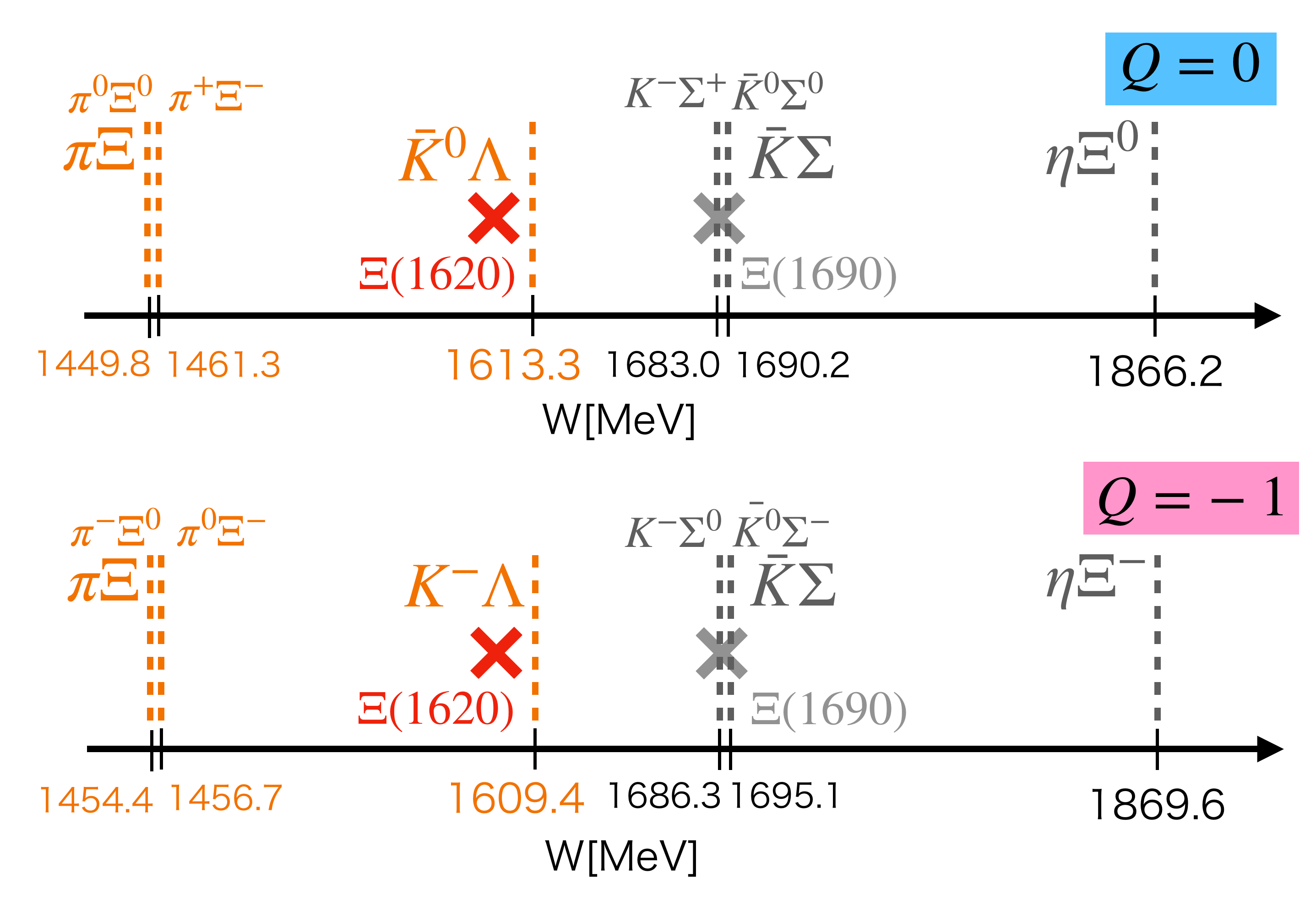}
    \caption{The threshold energies of the meson-baryon channels with strangeness $-2$ and the $\Xi$ resonances. The upper (lower) panel shows the neutral channels with $Q=0$ (charged channels with $Q=-1$).}
  \label{fig:thresholdsq06}
\end{figure}

% this work
Given that new data have been obtained by Belle and ALICE, in this paper, we discuss the nature of $\Xi(1620)$ by focusing on the location of the resonance pole in relation with the $K^-\Lambda$ scattering length. For this purpose, we classify the eigenstates near the threshold with respect to the value of the scattering length based on the effective range expansion. We then construct models which reproduces the experimental results by using chiral unitary approach. First, we construct a model with the $\Xi(1620)$ quasibound state below the threshold having a narrow width as indicated by Belle. Second, we construct a model by reproducing the $K^{-}\Lambda$ scattering length obtained by ALICE. Note that there are charged and neutral meson-baryon channels for $\Xi(1620)$ as shown in Fig.~\ref{fig:thresholdsq06}. The $\pi^+\Sigma^+$ spectrum of Belle corresponds to the neutral channel while the $K^-\Lambda$ scattering length is in the charged channel.

% threshold effect
When we investigate the spectrum of $\Xi(1620)$, it is important to pay attention to the effect from the threshold. The peak of $\Xi(1620)$ appears near the $\bar{K}\Lambda$ threshold at $1613.3\ {\rm MeV}$. In general, it is known that the near threshold spectrum is affected by the existence of the threshold, so we need to consider the threshold effect on the spectrum of the $\Xi(1620)$ resonance. The threshold effects can be examined in the chiral unitary approach which contains the threshold dynamics in the coupled-channel meson-baryon scattering.

The outline of this paper is as follows. In Sec.~\ref{sec:formulation}, we introduce the chiral unitary approach. We then discuss the relation between the complex scattering length and the pole near the threshold through the effective range expansion in Sec.~\ref{sec:polescat}. In Sec.~\ref{sec:numericalresult}, we first construct a model which generates $\Xi(1620)$ slightly below the $\bar{K}\Lambda$ threshold as indicated by the Belle result~(Model 1). Next, using the $K^-\Lambda$ scattering length by ALICE, we construct another model~(Model 2) and search for the pole of the scattering amplitude. The classification scheme of Sec.~\ref{sec:polescat} is applied to the results of Model 1 and Model 2 to investigate the nature of $\Xi(1620)$. Finally, we compare Model 1 with Model 2 by considering the experimental uncertainties in order to explore the model which satisfies both the constrains. In the last section, we summarize this work and present future prospects. The preliminary results of Sec.~\ref{subsec:Model1} is reported in the proceedings of conference~\cite{Nishibuchi:2022zfo}.

\section{Formulation of chiral unitary approach}\label{sec:formulation}

In this section, we formulate the coupled channel scattering amplitude by the chiral unitary approach~\cite{Kaiser:1995eg,Oset:1997it,Oller:2000fj,Hyodo:2011ur}. In general, the scattering amplitude $T_{ij}$ satisfies the scattering equation 
\begin{align}\label{eq:lseqchiral1}
T_{ij}(W)=V_{ij}(W)+\sum_kV_{ik}(W)G_k(W)T_{kj}(W), 
\end{align}
which is written by the loop function $G_{k}(W)$ and the interaction kernel $V_{ik}(W)$. $W$ is the total energy of the system and the indices $i,j,k$ specify the coupled channel. The meson-baryon scattering channels related to the $\Xi$ resonance with strangeness $-2$ are $\pi^0\Xi^0,\pi^+\Xi^-,\bar{K}^0\Lambda,K^-\Sigma^+,\bar{K}^0\Sigma^0$ and $\eta\Xi^0$ in the neutral channels and $\pi^-\Xi^0,\pi^0\Xi^-,K^-\Lambda,K^-\Sigma^0,\bar{K}^0\Sigma^-$ and $\eta\Xi^-$ in the negatively charged channels (Fig.~\ref{fig:thresholdsq06}). The interaction kernel $V_{ij}(W)$ represents the interaction from $j$ to $i$ channel. In this work, we use the Weinberg-Tomozawa term
\begin{align} \label{eq:wtinteraction}
V_{ij}^{WT}(W)=-\frac{C_{ij}}{4f_if_j}(2W-M_i-M_j)\sqrt{\frac{M_i+E_i}{2M_i}}\sqrt{\frac{M_j+E_j}{2M_j}},
\end{align}
which is the leading order term in chiral perturbation theory, as an s-wave interaction which is dominant at low energy. $f_i$, $M_i$ and $E_i$ are the meson decay constant, baryon mass and energy of the baryon in channel $i$, respectively. $C_{ij}$ is the group theoretical factor expressing the nature of the interaction. Explicit values of $C_{ij}$ are shown in Table.~\ref{tab:Cneutral} and Table.~\ref{tab:Cnegative} for neutral and negatively charged channels, respectively.

The loop function $G_k(W)$ is given as follows:
\begin{align}\label{eq:loopfunc1}
G_k(W)=i\int\frac{d^4q}{(2\pi)^4}\frac{2M_k}{(P-q)^2-M_k^2+i0^+}\frac{1}{q^2-m_k^2+i0^+},
\end{align}
where $P^\mu=(W,\bm{0})$ is the total four-momentum of the center mass system and $m_i$ is the meson mass in channel $i$. Because the $q$ integration of the loop function~\eqref{eq:loopfunc1} diverges logarithmically, some kind of regularization is required. When we apply the dimensional regularization, the finite part of the loop function is obtained as
\begin{align}
\begin{split}\label{eq:loopfuncfinite}
&G_k[W;a_k(\mu_{\mathrm{reg}})]
\\=&\frac{2M_k}{16\pi^2}\biggl[a_k(\mu_{\mathrm{reg}})+\ln\frac{m_kM_k}{\mu^2_{reg}}+\frac{M_k^2-m_k^2}{2W^2}\ln\frac{M_k^2}{m_k^2}
\\&+\frac{\lambda^{1/2}}{2W^2}\Bigl\{\ln(W^2-m_k^2+M_k^2+\lambda^{1/2})
\\&+\ln(W^2+m_k^2-M_k^2+\lambda^{1/2})
\\&-\ln(-W^2+m_k^2-M_k^2+\lambda^{1/2})
\\&-\ln(-W^2-m_k^2+M_k^2+\lambda^{1/2})\Bigr\}\biggr], 
\end{split}
\\\lambda^{1/2}=&\sqrt{W^4+m_k^4+M_k^4-2W^2m_k^2-2m_k^2M_k^2-2M_k^2W^2}, \label{eq:loopfuncfinitelambda}
\end{align}
where $\mu_{\mathrm{reg}}$ is the regularization scale and $a_i(\mu_{\mathrm{reg}})$ is the subtraction constant of channel $i$.

The nonrelativistic scattering amplitude $F_{ij}(W)$ is obtained from $T_{ij}(W)$ in Eq.~\eqref{eq:lseqchiral1} as
\begin{align}
F_{ij}(W)&=-\frac{\sqrt{M_iM_j}}{4\pi W}T_{ij}(W). \label{eq:chiralf}
\end{align}
The imaginary part of the nonrelativistic scattering amplitude $F_{ij}(W)$ corresponds to the spectrum. The scattering length of channel $i$ is determined by the elastic scattering amplitude at the threshold energy as
\begin{align}\label{eq:detscatl}
a_{0,i}=-F_{ii}(W=m_i+M_i).
\end{align}

If a resonance state exists, the scattering amplitude has a pole at the eigenenrgy in the complex energy plane. To perform the analytic continuation of the scattering amplitude,  it is important to consider the Riemann sheets of the complex energy plane. For a single channel scattering, the energy corresponding to the upper half of the complex momentum plane is called the [t] sheet (first Riemann sheet), and that corresponding to the lower half is named the [b] sheet (second Riemann sheet)~\cite{Pearce:1988rk}. In the complex energy plane, there is a branch cut on the positive real energy axis with the branch point at the origin. In a general coupled channel scattering, because the momentum in each channel should be determined for a given energy $E$, there are two Riemann sheets ([t], [b]) in each coupled channel. Hence, for an $n$-channel scattering, there are in total $2^n$ Riemann sheets. We denote the choice of the Riemann sheets as [bbtt$\cdots$] in the ordering of the channels with lower threshold energy. Classification of the eigenstates is important because the effects on the scattering amplitude on the real axis depends on the position and the Riemann sheets of the pole of the eigenstate.

 \begin{table}%[H] add [H] placement to break table across pages
 \caption{Group theoretical factor $C_{ij}$ for neutral channels.\label{tab:Cneutral}}
 \begin{ruledtabular}
\begin{tabular}{c c c c c c c}
& $K^-\Sigma^+$ & $\bar{K}^0\Sigma^0$ & $\bar{K}^0\Lambda$ & $\pi^+\Xi^-$ & $\pi^0\Xi^0$ & $\eta\Xi^0$\\
\hline
$K^-\Sigma^+$ & $1 $ & $-\sqrt{2}$ & $0$ & $0$ & $-1/\sqrt{2}$ & $-\sqrt{3/2}$\\
$\bar{K}^0\Sigma^0$ & $-\sqrt{2}$ & $0$ & $0$ & $-1/\sqrt{2}$ & $-1/2$ & $\sqrt{3/4}$\\
$\bar{K}^0\Lambda$ & $0 $ & $0$ & $0$ & $-\sqrt{2/3}$ & $-1/2$ & $-3/2$\\
$\pi^+\Xi^-$ & $0$ & $-1\sqrt{2}$ & $-\sqrt{3/2}$ & $1$ & $-\sqrt{2}$ & $0$\\
$\pi^0\Xi^0$ & $-1/\sqrt{2}$ & $-1/2$ & $\sqrt{3/4}$ & $-\sqrt{2}$ & $0$ & $0$\\
$\eta\Xi^0$ & $-\sqrt{3/2}$ & $\sqrt{3/4}$ & $-3/2$ & $0$ & $0$ & $0$\\
\end{tabular}
 \end{ruledtabular}
 \end{table}

 \begin{table}%[H] add [H] placement to break table across pages
 \caption{Group theoretical factor $C_{ij}$ for negative charged channels.\label{tab:Cnegative}}
 \begin{ruledtabular}
\begin{tabular}{c c c c c c c}
& $\bar{K}^0\Sigma^-$ & $K^-\Sigma^0$ & $K^-\Lambda$ & $\pi^-\Xi^0 $ & $\pi^0\Xi^-$ & $\eta\Xi^-$ \\
 \hline 
$\bar{K}^0\Sigma^-$ & 1 & $\sqrt{2}$ & 0 & 0 & $1/\sqrt{2}$ & $-\sqrt{3/2}$ \\
$K^-\Sigma^0$ & $\sqrt{2}$ & 0 & 0 & $1/\sqrt{2}$ & $-1/2$ &$-\sqrt{3/4}$ \\
$K^-\Lambda$ & 0 & 0 & 0 & $-\sqrt{3/2}$ & $-\sqrt{3/4}$ & $-3/2$\\
$\pi^-\Xi^0$ & 0 & $1/\sqrt{2}$ & $-\sqrt{3/2}$ & 1 & $\sqrt{2}$ & 0 \\ 
$\pi^0\Xi^-$ & $1/\sqrt{2}$ & $-1/2$ & $-\sqrt{3/4}$ & $\sqrt{2}$ & 0 & 0 \\
$\eta\Xi^-$ & $-\sqrt{3/2}$ & $-\sqrt{3/4}$ & $-3/2$ & 0 & 0 & 0 \\
\end{tabular}
 \end{ruledtabular}
 \end{table}

\section{Scattering length and pole position}\label{sec:polescat}

Here we first summarize the basic properties of the Riemann sheets of the complex energy plane for the two-channel scattering, and classify the eigenstate represented by the pole of the scattering amplitude in Sec.~\ref{sec:energy}. We then consider the classification of the eigenstate in the corresponding complex momentum plane, in order to express it by the eigenmomentum in Sec.~\ref{sec:momentum}. Finally, we relate the eigenmomentum to the scattering length for the classification of the eigenstate in Sec.~\ref{sec:slength}.

\subsection{Eigenstates in complex energy plane}\label{sec:energy}

For a near-threshold resonance such as $\Xi(1620)$, it is expected that the scattering length is strongly affected by the presence of the pole of the resonance. In this section, we discuss the relation between the scattering length and the near-threshold pole in a simple two-channel scattering (see also Refs.~\cite{Pearce:1988rk,Yamada:2021azg}). Bearing in mind the $\pi\Xi$-$\bar{K}\Lambda$ system with $\Xi(1620)$ near the $\bar{K}\Lambda$ threshold, we consider the eigenstate near the higher energy threshold. Because the Riemann sheet is specified by the choice of the momentum of each channel, there are four sheets denoted by [tt], [tb], [bt], and [bb].

\begin{figure}[tbp]
  \begin{center}
    \includegraphics[width=8cm]{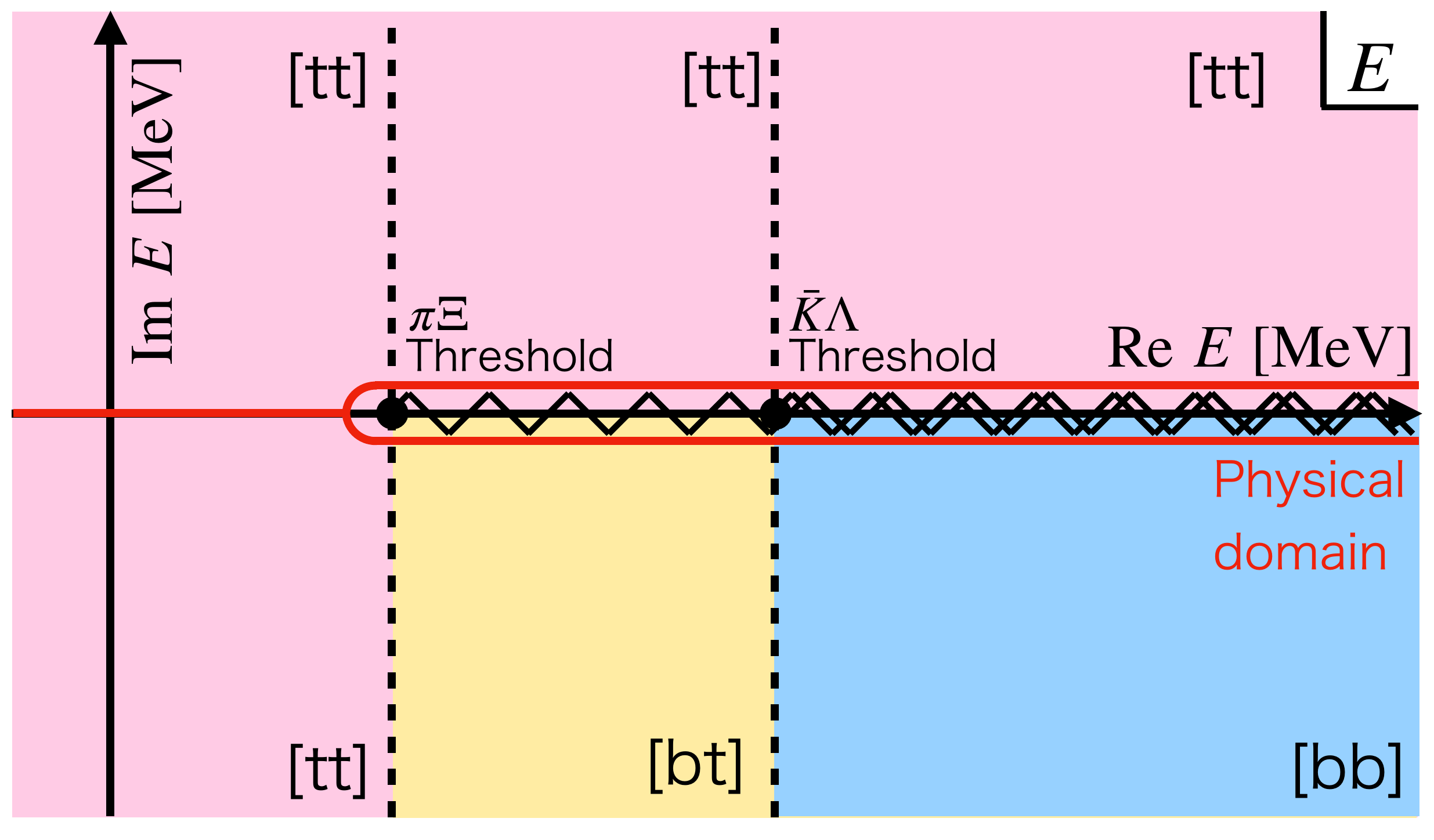}
    \caption{
    The physically relevant Riemann sheet of the complex energy plane for the $\pi\Xi$-$\bar{K}\Lambda$ scattering. 
    The solid line represents the physical domain.}
    \label{fig:riemannsheet2chpre}
  \end{center}
\end{figure}

Physical scattering occurs for a real and positive momentum (physical domain) in a single channel scattering. In the complex energy plane, the physical domain exists on the positive real axis between the [t] sheet with positive imaginary part ($E+i\epsilon$ with $\epsilon>0$) and the [b] sheet with negative imaginary part ($E-i\epsilon$). Below the threshold, the physical domain is analytically continued to the pure imaginary momentum with a positive imaginary part, corresponding to the real energy axis in the [t] sheet. Namely, physical domain is in the [t] sheet for $E+i\epsilon$ for all energy, while for $E-i\epsilon$ it is in the [t] sheet below the threshold and in the [b] sheet above the threshold. Generalizing this for the coupled-channel scattering, the physical domain is connected to the sheet with the choice of [t] for all channels ([tt$\cdots$] sheet) at $E+i\epsilon$ and the sheet with the choice of [b] ([t]) sheet for the open (closed) channels at $E-i\epsilon$. In the following, we call this the physically relevant Riemann sheet. The physically relevant Riemann sheet for the $\pi\Xi$-$\bar{K}\Lambda$ scattering is illustrated in Fig.~\ref{fig:riemannsheet2chpre}.

The eigenstate pole can be generated on the real energy axis below the $\pi\Xi$ threshold  and the lower half plane above the $\pi\Xi$ threshold. The pole below the $\pi\Xi$ threshold represents the stable bound state, and the pole above the $\pi\Xi$ threshold with finite imaginary part stands for the unstable resonance. In this work, we call the pole between the $\pi\Xi$ and $\bar{K}\Lambda$ thresholds the quasibound state ($QB$), because it can be interpreted as the bound state in the $\bar{K}\Lambda$ channel which acquires the decay width through the coupling to the $\pi\Xi$ continuum. Although the quasibound state can be regarded as a resonance in the $\pi\Xi$ scattering, here we use ``quasibound state'' in order to distinguish it from the pole above the $\bar{K}\Lambda$ threshold, which we call the resonance $R$. The pole of $QB$ locates on the [bt] sheet, which is different from the [bb] sheet where $R$ exists. In this way, because the Riemann sheet is different, $QB$ below the threshold is not continuously connected to $R$ above the threshold.

\begin{figure}[tbp]
  \begin{center}
    \includegraphics[width=8.5cm]{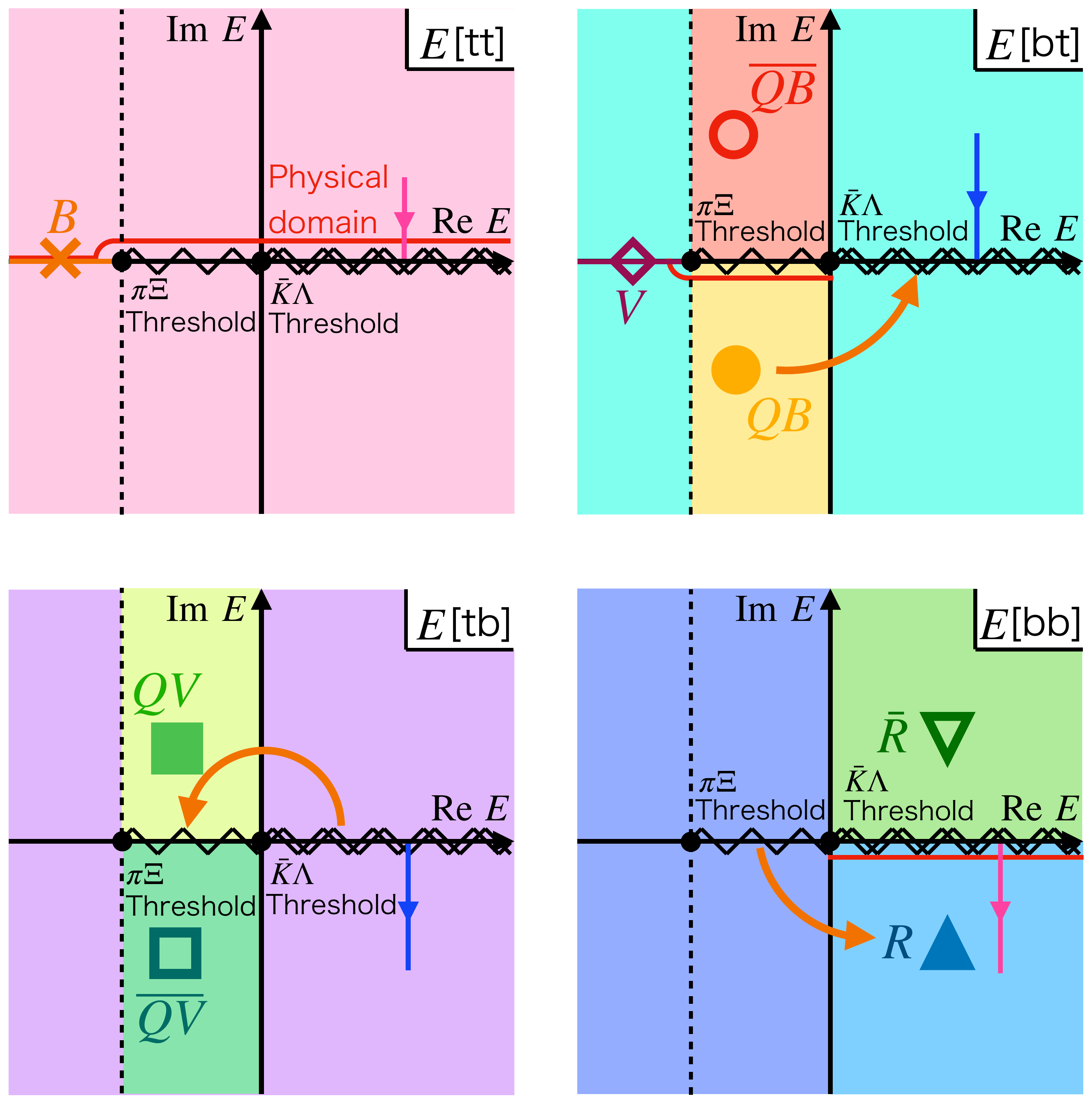}
    \caption{
    Riemann sheets of the complex energy plane for the $\pi\Xi$-$\bar{K}\Lambda$ scattering. 
    Poles are classified into $B$ (bound state, cross), $V$ (virtual state, diamond), $R$ (resonance, filled triangle), $\bar{R}$ (anti-resonance, open triangle), $QB$ (quasibound state, filled circle), $\overline{QB}$ (quasibound state, open circle), $QV$ (quasivirtual state, filled square), $\overline{QV}$ (anti-quasivirtual state, open square). 
    The solid line represents the physical domain.}
    \label{fig:riemannsheet2ches}
  \end{center}
\end{figure}

In Fig.~\ref{fig:riemannsheet2ches}, we show the Riemann sheets of the complex energy plane by setting the $\bar{K}\Lambda$ threshold at the origin. There are four Riemann sheets, [tt], [tb], [bt], and [bb] for the two-channel scattering. Above the $\pi\Xi$ threshold, the physical domain is in the [tt] sheet with $E+i\epsilon$, in the [bt] sheet with $E-i\epsilon$ between the $\pi\Xi$ and $\bar{K}\Lambda$ thresholds, and in the [bb] sheet with $E-i\epsilon$ above the $\bar{K}\Lambda$ threshold.

Next, we classify the eigenstates expressed by the pole of the scattering amplitude.
We define the resonance ($R$) as a pole above the $\bar{K}\Lambda$ threshold in the lower half plane of the [bb] sheet, and the quasibound state ($QB$) as a pole between the $\pi\Xi$ and $\bar{K}\Lambda$ thresholds in the lower half plane of the [bt] sheet. According to the Schwarz reflection principle, the existence of the $R$ and $QB$ pole in the lower half plane indicates a pole in the upper half plane at the position symmetric with respect to the real axis. The pole accompanied by the resonance is called the anti-resonance ($\bar{R}$), and here we call the one associated with the quasibound state the anti-quasibound state ($\overline{QB}$). While $R$ and $\bar{R}$ ($QB$ and $\overline{QB}$) exist at the same distance from the real axis, the pole of $R$ ($QB$) has stronger effect on the physical scattering because the physical domain is in the lower half plane (see Fig.~\ref{fig:riemannsheet2ches}). The stable bound state $(B)$ can be found on the real axis below the $\pi\Xi$ threshold in the [tt] sheet, and the virtual state is in the [bt] sheet.

Let us consider the behavior of the $QB$ pole by gradually adjusting the coupling to the $\pi\Xi$ channel. When the coupling to the $\pi\Xi$ channel is reduced, the width of a quasibound state $QB$ becomes smaller, and the pole approaches the real energy axis. In the zero coupling limit~\cite{Eden:1964zz}, the pole eventually reaches the real axis between the [tt] and [bt] sheets, which is nothing but a bound state in the single $\bar{K}\Lambda$ channel. In the same way, a virtual state in the single $\bar{K}\Lambda$ channel in the zero coupling limit should exist on the real energy axis between the [tb] and [bb] sheets. When the coupling to the $\pi\Xi$ channel is switched on, it is expected that the virtual state pole moves into the complex energy plane of the [tb] sheet\footnote{In principle, the virtual pole can move into the [bb] sheet, but as we shall show below, the pole on the [bb] sheet cannot be reproduced only by the scattering length. We therefore expect that the near-threshold quasivirtual pole should move into the [tb] sheet.}. We call this pole the ``quasivirtual state'' ($QV$), which is a would-be virtual state in the absence of the channel coupling. 

Some remarks on the quasivirtual state $QV$ are in order. First, because the pole of $QV$ is not on the physically relevant Riemann sheet shown in Fig.~\ref{fig:riemannsheet2chpre}, we do not expect the peak structure in the corresponding spectrum on the real axis. Nevertheless, the existence of a quasivirtual pole near the threshold implies a strong cusp effect at the $\bar{K}\Lambda$ threshold~\cite{Yamada:2021azg}. Second, when we consider the transition from the quasibound state $QB$ to the resonance $R$, the expected pole trajectory 
goes through the quasivirtual state $QV$ as indicated by the arrows in Fig.~\ref{fig:riemannsheet2ches} (see Ref.~\cite{Pearce:1988rk}). This is analogous to the transition from the bound state to the resonance through the virtual state in the single-channel scattering~\cite{Hyodo:2013iga,Hyodo:2014bda}. Third, the imaginary part of the $QV$ pole is positive, in contrast to $QB$ and $R$ which have a negative imaginary part. From the discussion of the pole trajectory, it is clear that the pole with a positive imaginary part is closer to the physical domain than that with a negative imaginary part. 

\subsection{Eigenstates in complex momentum plane}\label{sec:momentum}

\begin{figure}[tbp]
  \begin{center}
    \includegraphics[width=8.5cm]{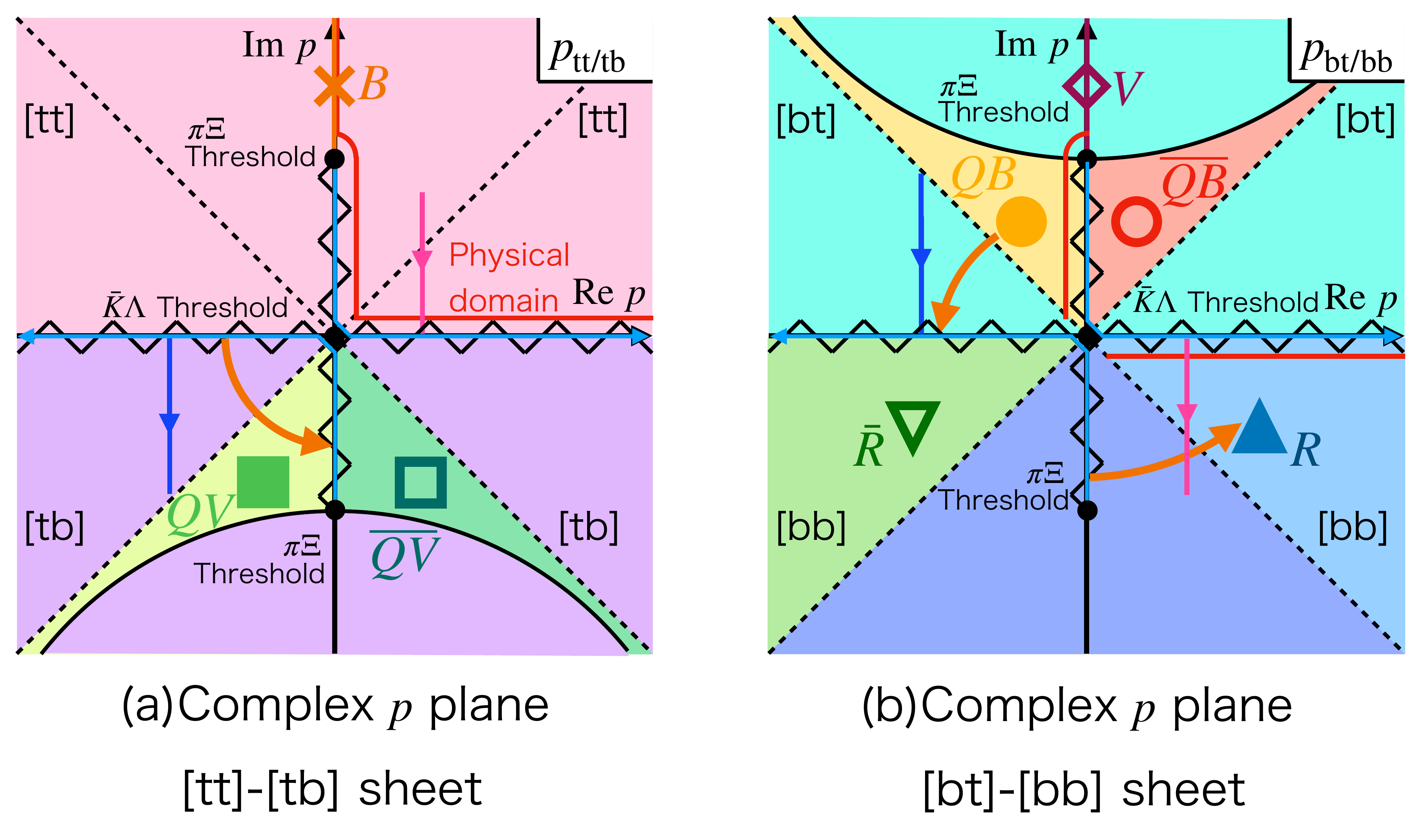}
    \caption{
    Riemann sheets of the complex momentum plane of the $\bar{K}\Lambda$ channel for the $\pi\Xi$-$\bar{K}\Lambda$ scattering. 
    (a): $p_{tt/tb}$ sheet, (b): $p_{bt/bb}$ sheet.
    Poles are classified into $B$ (bound state, cross), $V$ (virtual state, diamond), $R$ (resonance, filled triangle), $\bar{R}$ (anti-resonance, open triangle), $QB$ (quasibound state, filled circle), $\overline{QB}$ (quasibound state, open circle), $QV$ (quasivirtual state, filled square), $\overline{QV}$ (anti-quasivirtual state, open square). The solid line represents the physical domain.}
    \label{fig:riemannsheet2chp1}
  \end{center}
\end{figure}

Next, we consider the same classification of the eigenstate pole in the complex momentum plane. The complex momentum plane of the $\bar{K}\Lambda$ channel has two Riemann sheets to specify the sign of the $\pi\Xi$ momentum. In Fig.~\ref{fig:riemannsheet2chp1} we show two complex momentum plane of the $\bar{K}\Lambda$ channel. In panel (a), the Riemann sheet of the $\pi\Xi$ channel is fixed to the [t] sheet, and (b) to the [b] sheet. The upper half plane of Fig.~\ref{fig:riemannsheet2chp1} (a) corresponds to the [tt] sheet of the complex energy plane, and the lower half plane to the [tb] sheet. We therefore call (a) the complex $p_{\mathrm{tt/tb}}$ plane and (b) the $p_{\mathrm{bt/bb}}$ plane.

\begin{figure}[tbp]
  \begin{center}
    \includegraphics[width=8.5cm]{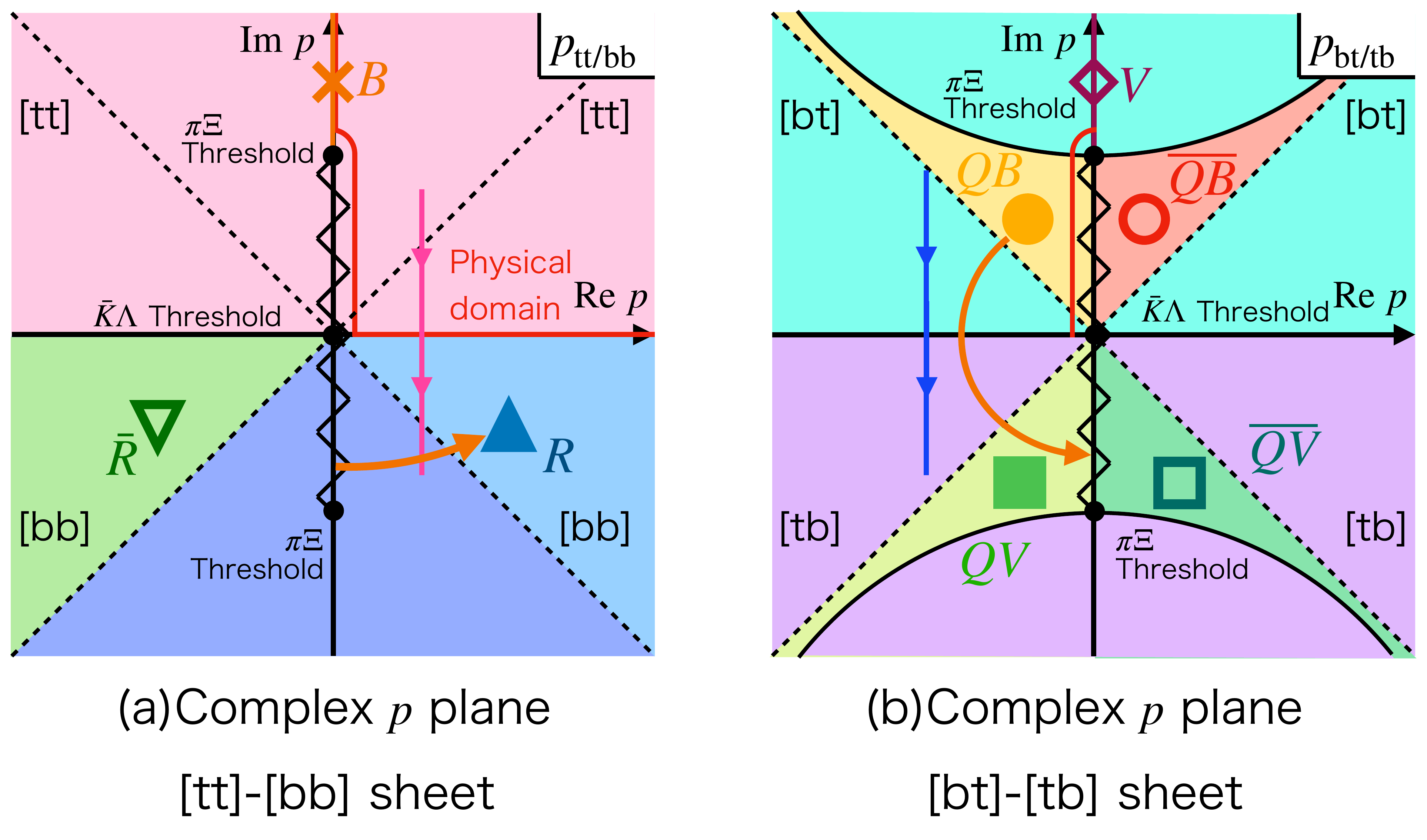}
    \caption{    
    Riemann sheets of the complex momentum plane of the $\bar{K}\Lambda$ channel for the $\pi\Xi$-$\bar{K}\Lambda$ scattering. 
    (a): $p_{tt/bb}$ sheet, (b): $p_{bt/bb}$ sheet.
    Poles are classified into $B$ (bound state, cross), $V$ (virtual state, diamond), $R$ (resonance, filled triangle), $\bar{R}$ (anti-resonance, open triangle), $QB$ (quasibound state, filled circle), $\overline{QB}$ (quasibound state, open circle), $QV$ (quasivirtual state, filled square), $\overline{QV}$ (anti-quasivirtual state, open square). The solid line represents the physical domain.}
    \label{fig:riemannsheet2chp2}
  \end{center}
\end{figure}

By measuring the energy $E$ from the $\bar{K}\Lambda$ threshold, the $\pi\Xi$ threshold energy is negative, $E=-\Delta$ with $\Delta>0$. Since the relation between the $\bar{K}\Lambda$ momentum $p$ and the energy $E$ is $p=\pm\sqrt{2\mu E}$, the pure imaginary momenta $p=\pm i\sqrt{2\mu\Delta}$ correspond to the $\pi\Xi$ threshold, as seen in Fig.~\ref{fig:riemannsheet2chp1}. In the complex energy plane, the branch cut runs from the threshold on the positive real energy axis. In the complex momentum plane in Fig.~\ref{fig:riemannsheet2chp1}, the branch cuts run also from the threshold momenta to $\pm\infty$ on the real axis. The cut on the real axis in Fig.~\ref{fig:riemannsheet2chp1} (a) indicates that the [tt] sheet is not directly connected to the [tb] sheet (see also Fig.~\ref{fig:riemannsheet2ches}). In the same way, the [bt] sheet is separated from the [bb] sheet by the branch cut in panel (b). The cut on the real axis in Fig.~\ref{fig:riemannsheet2chp1} can be eliminated by exchanging the lower half plane, as shown in Fig.~\ref{fig:riemannsheet2chp2}. The complex $p_{\mathrm{tt/bb}}$ plane consists of the [tt] sheet in the upper half plane and the [bb] sheet in the lower half plane, and $p_{\mathrm{bt/tb}}$ of the [bt] sheet in the upper half plane and the [tb] sheet in the lower half plane. In this case, the upper and lower half planes are smoothly connected without the cut.

We then classify the eigenstates in the complex momentum plane. The poles of the quasibound state $QB$ and $\overline{QB}$ appear in the upper half plane of the $p_{\mathrm{bt/tb}}$ sheet in Fig.~\ref{fig:riemannsheet2chp2}. By denoting the phases of the complex momentum $p$ and the energy $E$ with respect to the $\bar{K}\Lambda$ threshold as
\begin{align}
p=|p|e^{i\theta_p},\ E=|E|e^{i\theta_E},
\end{align}
the relation $E=p^{2}/(2\mu)$ leads to 
\begin{align}\label{eq:thetqepco}
2\theta_p=\theta_E,
\end{align}
where $\nu$ is the reduced mass of the $\bar{K}\Lambda$ channel. From the [bt] sheet of the complex energy plane in Fig.~\ref{fig:riemannsheet2ches}, the phase of the energy $\theta_{E}$ of $QB$ is restricted as $\pi\leq \theta_E\leq3\pi/2$. Equation~\eqref{eq:thetqepco} then gives the corresponding phase of the momentum as
\begin{align}\label{eq:krangethetaqb}
QB:\quad \frac{\pi}{2}&\leq\theta_p\leq\frac{3}{4}\pi\ [\mathrm{bt}].
\end{align}
In addition, because the $QB$ pole should appear above the $\pi\Xi$ threshold energy $E=-\Delta$, the condition $-\Delta\leq\mathrm{Re}\ E$ holds. Denoting the complex momentum as $p=\gamma+i\kappa$ with $\gamma,\kappa\in \mathbb{R}$, this condition can be expressed as
\begin{align}
-\sqrt{\gamma^2+2\mu\Delta}\leq\kappa\leq\sqrt{\gamma^2+2\mu\Delta}.
\end{align}
Namely, the $QB$ pole can be found in the $p_{\mathrm{bt/bb}}$ plane below the hyperbola $\kappa=\sqrt{\gamma^2+2\mu\Delta}$ with \eqref{eq:krangethetaqb}, as shown in Fig.~\ref{fig:riemannsheet2chp2} (b). The allowed region for the pole of $\overline{QB}$ should be symmetric with respect to the imaginary axis, as shown in Fig.~\ref{fig:riemannsheet2chp2} (b).

For the resonance $R$, the pole should appear above the $\bar{K}\Lambda$ threshold and its phase of the complex energy satisfies
\begin{align}
\frac{7}{2}\pi\leq\theta_E\leq4\pi.
\end{align}
Corresponding phase of the momentum is restricted as
\begin{align}\label{eq:krangetheta1}
R: \frac{7}{4}\pi \leq\theta_p\leq2\pi\quad [\mathrm{bb}].
\end{align}
The pole region for the quasivirtual state $QV$ is obtained in the same manner with the quasibound state $QB$. The allowed phase of the eigenenergy is
\begin{align}
\frac{5}{2}\pi\leq\theta_E\leq 3\pi,
\end{align}
which leads to
\begin{align}\label{eq:krangethetaqv}
QV: \frac{5}{4}\pi \leq\theta_p\leq \frac{3}{2}\pi\quad [\mathrm{tb}],
\end{align}
and the condition $-\Delta<E$ is given by
\begin{align}
-\sqrt{\gamma^2+2\mu\Delta}\leq\kappa .
\end{align}

\subsection{Relation with scattering length}\label{sec:slength}

Now we are in a position to relate the eigenmomntum with the scattering length. The scattering length of the higher energy ($\bar{K}\Lambda$) channel $a_{0}$ is determined by the elastic scattering amplitude at the threshold as in Eq.~\eqref{eq:detscatl}. In contrast to the single-channel scattering where the scattering length $a_{0}$ is real, the scattering length has an imaginary part when the coupling to the lower energy decay channel is switched on. Note that the sign of the imaginary part of $a_{0}$ is shown to be negative due to the optical theorem. 

When the magnitude of the scattering length $|a_{0}|$ is sufficiently large, the near-threshold scattering amplitude can be approximated as $f(p)=(-1/a_0-ip)^{-1}$, and the pole of the scattering amplitude is given by 
\begin{align}\label{eq:kinji}
p=\frac{i}{a_0}.
\end{align}
To relate this with the classification of the eigenstate, we define the phase of the scattering length $\theta_{a_{0}}$ as 
\begin{align}
\begin{split}
a_0&=|a_0|e^{i\theta_{a_0}},
\end{split}
\end{align}
From Eq.~\eqref{eq:kinji}, the relation between $\theta_{a_{0}}$ and $\theta_{p}$ is given by
\begin{align}
\begin{split}
\theta_{a_0}=\frac{\pi}{2}-\theta_p, \label{argka}
\end{split}
\end{align}
Thus, the conditions in Eqs.~\eqref{eq:krangethetaqb}, \eqref{eq:krangetheta1} and \eqref{eq:krangethetaqv} are translated into those for $\theta_{a_{0}}$ as
\begin{align}
\begin{split}
QB:\frac{7\pi}{4}\leq\theta_{a_0}\leq2\pi, \\
R: \frac{\pi}{2} \leq\theta_{a_0}\leq\frac{3\pi}{4} , \\
QV: \pi\leq\theta_{a_0}\leq \frac{5\pi}{4}. 
\end{split}
\end{align}
Namely, when the scattering amplitude has the quasibound state $QB$, the real part of the scattering length is positive and the imaginary part is negative with a smaller magnitude than the real part. For the resonance $R$, the imaginary part is positive and the real part is negative with a smaller magnitude than the imaginary part. For the quasivirtual state $QV$, both the real and imaginary parts is negative with the real part having the larger magnitude than the imaginary part. These can be expressed as
\begin{align}
\begin{split}
QB:&\ \mathrm{Re}\ a_0>0,\quad \mathrm{Im}\ a_0<0, \quad |\mathrm{Im}\ a_0|<\mathrm{Re}\ a_0,
\\R:&\ \mathrm{Re}\ a_0<0,\quad \mathrm{Im}\ a_0>0, \quad |\mathrm{Re}\ a_0|<\mathrm{Im}\ a_0.
\\QV:&\ \mathrm{Re}\ a_0<0,\quad \mathrm{Im}\ a_0<0, \quad |\mathrm{Im}\ a_0|<|\mathrm{Re}\ a_0|.
\end{split}
\end{align}
The result for $R$ however shows the positive imaginary part which contradicts with the optical theorem. This indicates that the resonance $R$ above the $\bar{K}\Lambda$ threshold cannot be described by the approximation $f(p)=(-1/a_{0}-ip)^{-1}$, and the contribution from the effective range is needed.

In general, the phase of the scattering length $\theta_{a_{0}}$ is related to the Riemann sheet as 
\begin{align}
\begin{split}
\frac{\pi}{2} \leq \theta_{a_0} \leq \pi &:\ [\mathrm{bb}],
\\\pi \leq \theta_{a_0} \leq \frac{3}{2}\pi &:\ [\mathrm{tb}],
\\\frac{3\pi}{2}\leq\theta_{a_0}\leq 2\pi &:\ [\mathrm{bt}],
\end{split}
\end{align}
which can be expressed by the real and imaginary parts of the scattering length as 
\begin{align}
\begin{split}
&\ \mathrm{Re}\ a_0<0,\quad \mathrm{Im}\ a_0>0 : \ [\mathrm{bb}],
\\ &\ \mathrm{Re}\ a_0<0, \quad \mathrm{Im}\ a_0<0 : \ [\mathrm{tb}],
\\ &\ \mathrm{Re}\ a_0>0, \quad \mathrm{Im}\ a_0<0 : \  [\mathrm{bt}].
\end{split}
\end{align}
As mentioned above, the scattering length for the pole in the [bb] sheet does not satisfy the optical theorem, and therefore such pole cannot be represented by the scattering amplitude approximated by only the scattering length.

When the effective range $r_{e}$ is included, the inverse scattering amplitude can be expanded by neglecting the $\mathcal{O}(p^4)$ terms as 
\begin{align}
f(p)=\frac{1}{-\frac{1}{a_0}+\frac{r_e}{2}p^2-ip},
\end{align}
where the pole of the scattering amplitude is obtained as
\begin{align}
\begin{split}
p&=\frac{i\pm\sqrt{-1+\frac{2r_e}{a_0}}}{r_e}.
\end{split}
\end{align}
In this case, the resonance pole $R$ in the [bb] sheet can be realized thanks to the contribution from the effective range. In other words, the description of $R$ requires the contribution of the effective range $r_{e}$ in addition to the scattering length $a_{0}$. In fact, it is known for the single channel scattering that the effective range with a large magnitude is needed to obtain a narrow resonance above the threshold~\cite{Hyodo:2013iga}. 

\section{Numerical results}\label{sec:numericalresult}

\subsection{Model 1: narrow width $\Xi(1620)$}\label{subsec:Model1}

In this section, based on the previous work of the chiral unitary approach~\cite{Ramos:2002xh}, we construct a model which generates the $\Xi(1620)$ resonance in accordance with the Belle result. The reported values of the mass $M_{R}$ and decay width $\Gamma_{R}$ of $\Xi(1620)$ are as follows~\cite{Belle:2018lws}:
\begin{align}\label{eq:Belleresult}
\begin{split}
M_R&=1610.4\pm6.0({\rm{stat.}})^{+6.1}_{-4.2}({\rm{syst.}})\ {\rm{MeV}}, 
\\\Gamma_R&=59.9\pm4.8({\rm{stat.}})^{+2.8}_{-7.1}({\rm{syst.}})\ {\rm{MeV}}.
\end{split}
\end{align}
The pole position $z_{\rm ex}$ corresponding to the central values of Eq.~\eqref{eq:Belleresult} is determined as 
\begin{align}
z_{\rm{ex}}=1610-30i\ {\rm{MeV}}, \label{eq:bellepole}
\end{align}
which is lower than the $\bar{K}^0\Lambda$ threshold at $1613.3$ MeV. In this case, therefore, $\Xi(1620)$ is classified as the quasibound state $QB$. We aim at constructing the model with a pole at $z_{\rm ex}$, starting from Set 1 in Ref.~\cite{Ramos:2002xh} which has a pole at $1607-140i$ MeV.

Numerical calculations are performed in the physical basis for the neutral channels with $Q=0$ because the Belle result has been obtained from the $\pi^{+}\Xi^{-}$ spectrum~\cite{Belle:2018lws}. In this case, there are six meson-baryon channels as shown in Fig.~\ref{fig:thresholdsq06}. The masses of hadrons are taken from Particle Data Group (PDG)~\cite{ParticleDataGroup:2022pth}, including the isospin symmetry breaking effect. We set the meson decay constants as $f_\pi=f_K=f_\eta=104.439$ MeV according to Ref.~\cite{Ramos:2002xh}.

\begin{table}[tbp]
\centering
\caption{
Subtraction constants of Model 1, Model 2, and Set 1 in Ref.~\cite{Ramos:2002xh}.
}
\label{tab:subtractioinconstants}
\begin{ruledtabular}
\begin{tabular}{c c c c c}
& $a_{\pi\Xi}$ & $a_{\bar{K}\Lambda}$ & $a_{\bar{K}\Sigma}$ & $a_{\eta\Xi}$ \\
\hline
Ref.~\cite{Ramos:2002xh} Set 1 & $-2.00$ & $-2.00$ & $-2.00$ & $-2.00$ \\
Model 1  & $-4.26$ & $-0.12$ & $-2.00$ & $-2.00$\\
%Model 1  & $-4.19$ & $-0.14$ & $-2.00$ & $-2.00$\\
Model 2& $-2.90$ & $\phantom{-}0.36$ & $-2.00$ & $-2.00$
\end{tabular}
\end{ruledtabular}
\end{table}

The subtraction constant $a_{i}$ can be used to adjust the model to reproduce the pole position~\eqref{eq:bellepole}. Here we do not consider the isospin symmetry breaking in $a_{i}$ and treat four constants shown in Table~\ref{tab:subtractioinconstants}. By examining the dependence of the $\Xi(1620)$ pole position on the subtraction constants, we find that the pole depends strongly on the subtraction constants in the $\pi\Xi$ and $\bar{K}\Lambda$ channels, while it is insensitive to those in the $\bar{K}\Sigma$ and $\eta\Xi$ channels. This can be understood from the fact that the energy of the $\Xi(1620)$ pole is close to the $\pi\Xi$ and $\bar{K}\Lambda$ thresholds as shown in Fig.~\ref{fig:thresholdsq06}. Hence, in this work, we fix $a_{\bar{K}\Sigma}=a_{\eta\Xi}=-2$ as in Ref.~\cite{Ramos:2002xh}, and adjust $a_{\pi\Xi}$ and $a_{\bar{K}\Lambda}$ to reproduce the pole position in Eq.~\eqref{eq:bellepole}.

\begin{figure}[tbp]
    \centering
    \includegraphics[width=8.5cm]{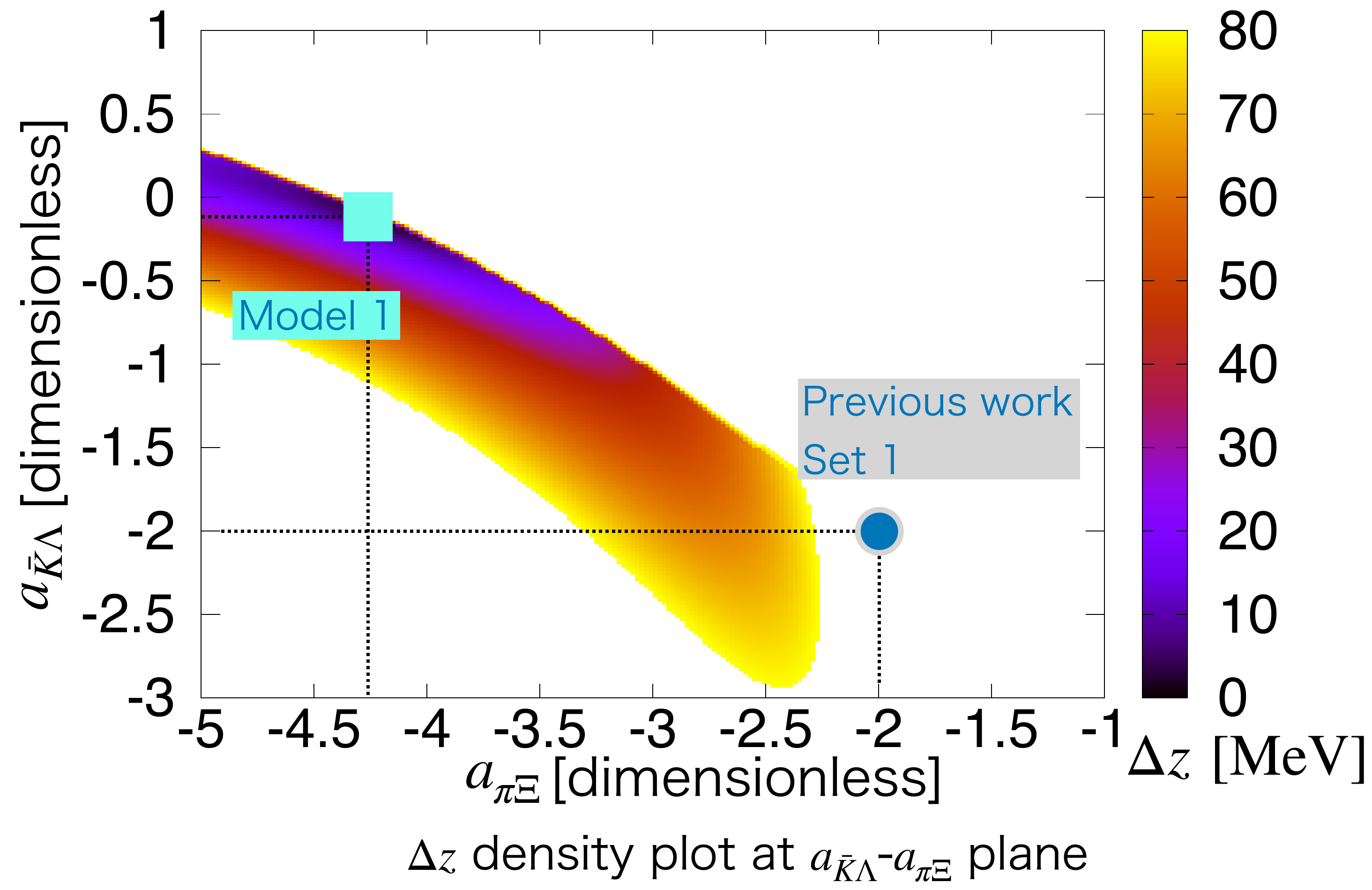}
    \caption{
    Density plot of $\Delta z$ in the $a_{\pi\Xi}$-$a_{\bar{K}\Lambda}$ plane.
    Square (circle) denotes the parameters of Model 1 (Set 1 in Ref.~\cite{Ramos:2002xh}).
    }
  \label{fig:deltazdensityplot}
\end{figure}

Denoting the pole position in the theoretical model as $z_{\rm th}$, we define the distance of $z_{\rm th}$ from Eq.~\eqref{eq:bellepole} in the complex energy plane as 
\begin{align*}
\Delta z=|z_{\rm{th}}-z_{\rm{ex}}| .
\end{align*}
Because $z_{\mathrm{th}}$ depends on the subtraction constants, we search for $(a_{\pi\Xi},a_{\bar{K}\Lambda})$ which minimize $\Delta z$. In Fig.~\ref{fig:deltazdensityplot}, we show the density plot of $\Delta z$ in the $a_{\pi\Xi}$-$a_{\bar{K}\Lambda}$ plane. $\Delta z$ can be minimized with $a_{\pi\Xi}=-4.26$ and $a_{\bar{K}\Lambda}=-0.12$, giving the pole position at 
\begin{align}
z_{\rm{th}}=1610-30i \ {\rm{MeV}}, \label{eq:model1polestat}
\end{align}
which reproduces Eq.~\eqref{eq:bellepole} in a 1 MeV precision. In the following, this parameter set is called Model 1. The subtraction constants are summarized in Table~\ref{tab:subtractioinconstants} together with those of Set 1 of Ref.~\cite{Ramos:2002xh}.

\begin{figure}[tbp]
\centering
\includegraphics[width=8cm]{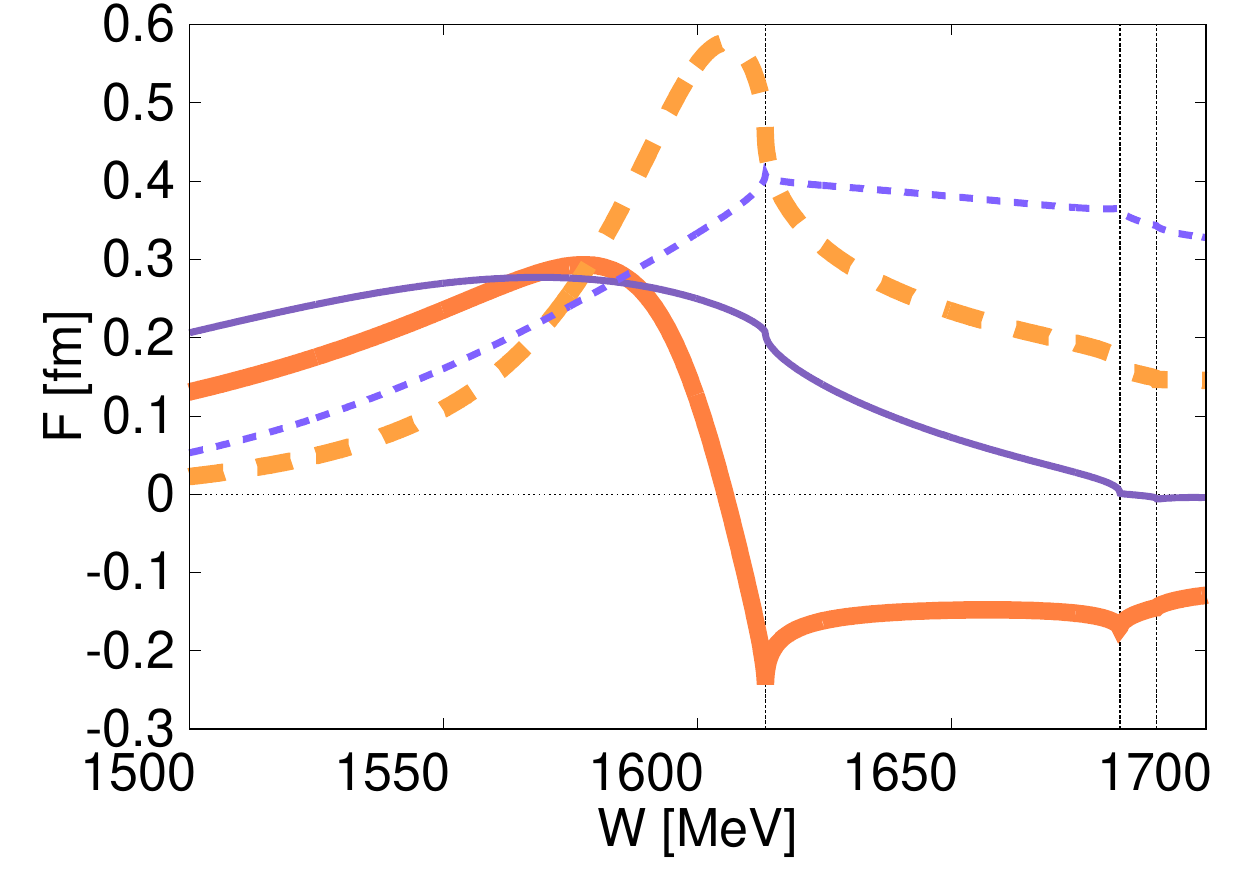}
\caption{
Comparison of the $\pi^+\Xi^-$ elastic scattering amplitude of Model 1 (thick lines) with the Set 1 in Ref.~\cite{Ramos:2002xh} (thin lines).
Solid (dashed) lines stand for the real (imaginary) parts.
Vertical dotted lines represent the thresholds of $\bar{K}^0\Lambda$, $K^-\Sigma^+$, $\bar{K}^0\Sigma^0$ from left to right.}
\label{fig:compprevandmodel1}
\end{figure}

To examine the peak structure of the spectrum on the real axis, we plot the scattering amplitude in Fig.~\ref{fig:compprevandmodel1} in comparison with Set 1 of Ref.~\cite{Ramos:2002xh}. The real and imaginary parts of the elastic $\pi^+\Xi^-$ scattering amplitude $F(W)$ defined in Eq.~\eqref{eq:chiralf} are shown as functions of the total energy $W$. The imaginary part of the amplitude of Ref.~\cite{Ramos:2002xh} (thin dashed line) does not show a prominent peak structure at the real part of the pole position (1607 MeV). This is because the large magnitude of the imaginary part ($-140i$ MeV) suppresses the effect of the pole on the real energy axis. In contrast, the result of Model 1 (thick lines) shows a typical resonance behavior of the scattering amplitude; the peak of the imaginary part and the zero of the real part at around $W=\text{Re } z_{\rm th}=1610$ MeV. In this way, we construct the model of chiral unitary approach which generates the $\Xi(1620)$ quasibound state with a relatively narrow width, as indicated by the Belle result.

\begin{figure}[tbp]
\centering
\includegraphics[width=8cm]{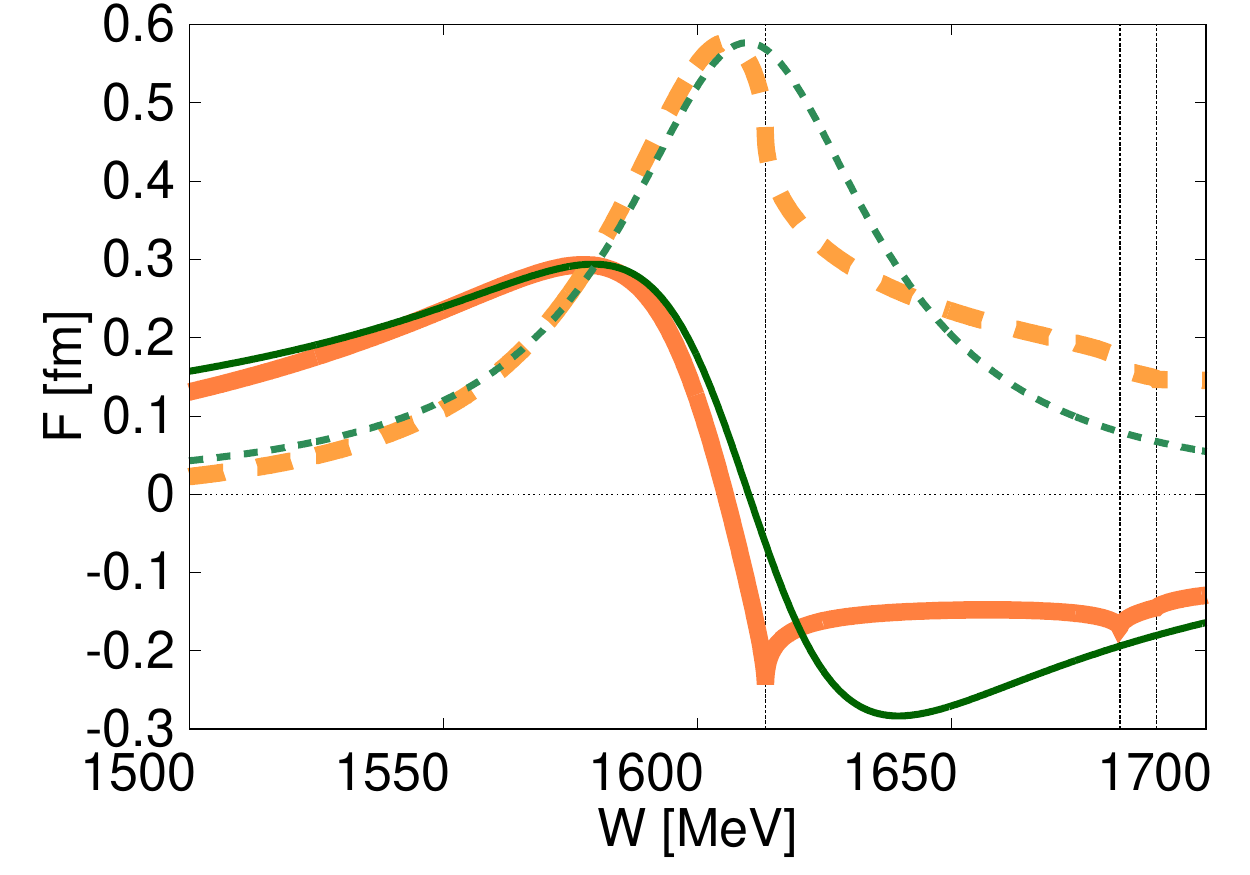}
\caption{
Comparison of the $\pi^+\Xi^-$ elastic scattering amplitude of Model 1 (thick lines) with the Breit-Wigner amplitude having the pole at the same position (thin lines).
Solid (dashed) lines stand for the real (imaginary) parts.
Vertical dotted lines represent the thresholds of $\bar{K}^0\Lambda$, $K^-\Sigma^+$, $\bar{K}^0\Sigma^0$ from left to right.}
\label{fig:compbreitandmodel1}
\end{figure}

It should however be noted that the position of the $\Xi(1620)$ pole is very close to the $\bar{K}^0\Lambda$ threshold at $1613.3\ \mathrm{MeV}$. In general, the peak near the threshold is known to be distorted by the effect of the threshold. To examine the threshold effect in Model 1, we plot the scattering amplitude of Model 1 together with the Breit-Wigner amplitude having a pole at the same position in Fig.~\ref{fig:compbreitandmodel1}. From the comparison of the imaginary parts in Fig.~\ref{fig:compbreitandmodel1}, we find that the peak of Model 1 is distorted by the presence of the $\bar{K}^0\Lambda$ threshold. In fact, the peak of Model 1 is at 1606 MeV, which is shifted about 4 MeV downward from the real part of the pole position (1610 MeV). In this way, we quantitatively discuss the $\bar{K}\Lambda$ threshold effect on the $\Xi(1620)$. This result indicates the importance of the threshold effect for the discussion on the mass and width of the near-threshold quasibound state.

So far we have discussed the model of $\Xi(1620)$ in the neutral meson-baryon channels where the pole (1610 MeV) is slightly below the $\bar{K}^0\Lambda$ threshold (1613.3 MeV). It is worth studying the $\Xi(1620)$ state in the negative charge channel where the threshold of the $K^-\Lambda$ is at 1609.4 MeV, which is lower than the pole of the neutral channel. We calculate the $Q=-1$ scattering amplitude with the $C_{ij}$ coefficients in Table~\ref{tab:Cnegative} and the corresponding hadron masses, using the subtraction constants of Model 1. The pole of the scattering amplitude is obtained at 
\begin{align}
z_{\rm th}=1609-30i\ \mathrm{MeV},
\quad (Q=-1) \label{eq:model1poleQm1}
\end{align}
which is lower than the $K^-\Lambda$ threshold. Namely, the negatively charged $\Xi(1620)$ is obtained as a quasibound state, as in the $Q=0$ channel. 

\subsection{Model 2: $K^-\Lambda$ scattering length} \label{subsec:Model2}

The ALICE experiment determines the $K^-\Lambda$ scattering length from the measurement of the correlation functions in the high-energy heavy ion collisions. The scattering length is obtained as~\cite{ALICE:2020wvi}
\begin{align}\label{eq:resultalice5}
\begin{split}
{\rm{Re}}\ f_0&=0.27\pm0.12({\rm{stat.}})\pm0.07({\rm{syst.}})\ {\rm{fm}},
\\{\rm{Im}}\ f_0&=0.40\pm0.11({\rm{stat.}})\pm0.07({\rm{syst.}})\ {\rm{fm}},
\end{split}
\end{align}
where $f_{0}$ is related to the scattering length in Eq.~\eqref{eq:detscatl} as $a_0=-f_0$. In general, because the scattering length determines the real and imaginary part of the scattering amplitude directly, it gives a stronger constraint on theoretical models than the fit to the spectrum. In the numerical calculation, we again adopt the physical basis with $Q=-1$ which contains the $K^-\Lambda$ channel as shown in Fig.~\ref{fig:thresholdsq06}. As before, isospin symmetric subtraction constants are used. 

First, we compare the scattering lengths in Set 1 of Ref.~\cite{Ramos:2002xh} and Model 1 with the result of the ALICE data. The central value of the ALICE measurement in the convention of $a_{0}$ is given by
\begin{align}
a_{0,\rm{ex}}=-0.27-0.40i \ {\rm{fm}}. \label{eq:fex}
\end{align}
The result of the $K^{-}\Lambda$ scattering length of Set 1 is $a_0=0.07-0.21i$ fm and that of Model 1 is $a_0=0.80-0.92i$ fm, in disagreement with the ALICE result in Eq.~\ref{eq:fex}.

\begin{figure}[tbp]
    \centering
    \includegraphics[width=8cm]{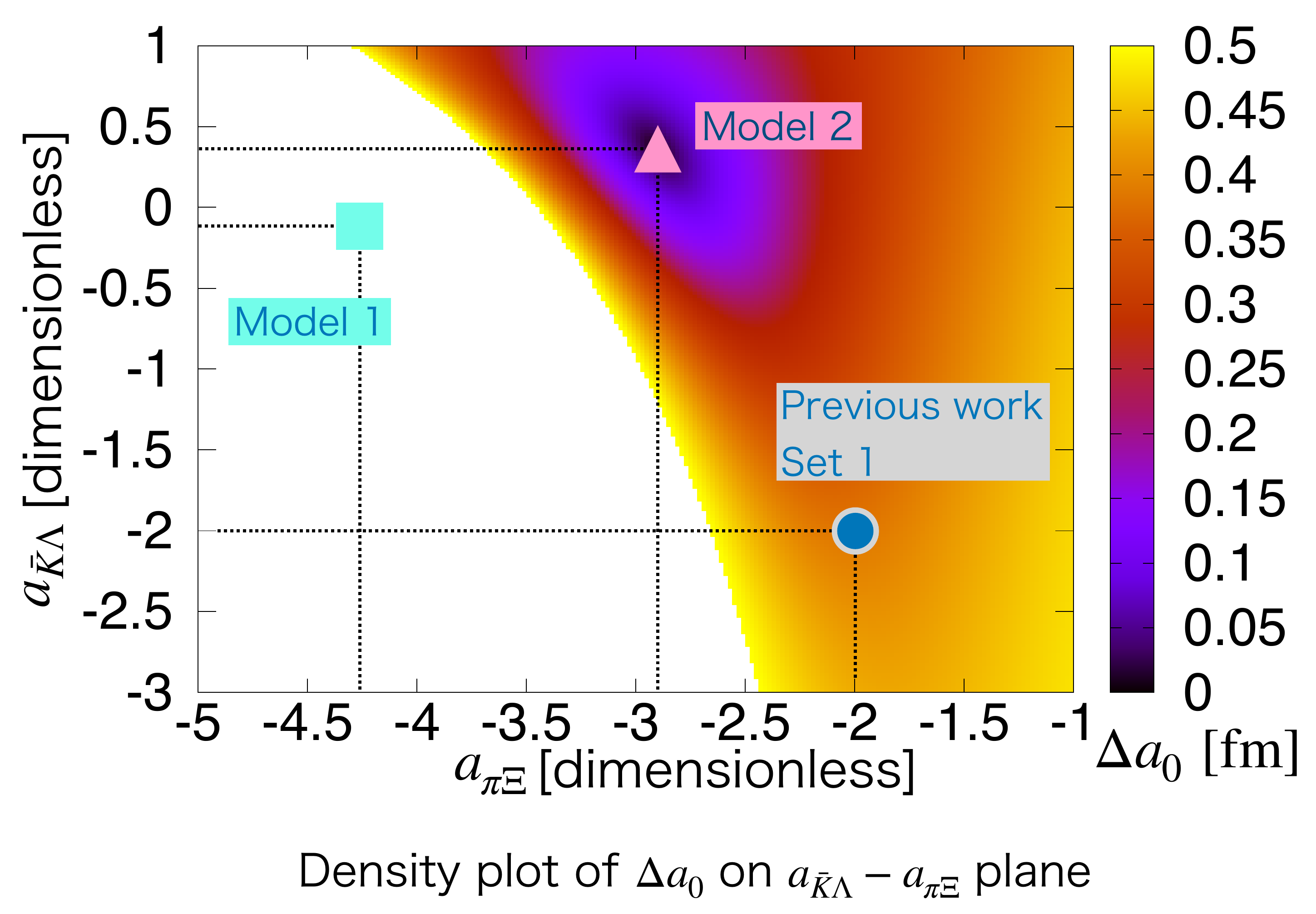}
    \caption{
    Density plot of $\Delta a_{0}$ in the $a_{\pi\Xi}$-$a_{\bar{K}\Lambda}$ plane. 
    Triangle, square, and circle denote the parameters of Model 2, Model 1, and Set 1 in Ref.~\cite{Ramos:2002xh}.
    }
  \label{fig:deltafdensityplot}
\end{figure}

Here we construct a model which reproduce Eq.~\ref{eq:fex}. As in the previous section, we optimize the subtraction constants in the $\pi\Xi$ and $\bar{K}\Lambda$ channels with $a_{\bar{K}\Sigma}=a_{\eta\Xi}=-2$ to reproduce the scattering length. Denoting the scattering length in the theoretical model as $a_{0,\rm{th}}$, we define its  deviation from the experimental data of $a_{0,\rm{ex}}$ in Eq.~\eqref{eq:fex} as 
\begin{align}
\Delta a_0=|a_{0,\rm{th}}-a_{0,\rm{ex}}|.
\end{align}
To search for the set of $a_{\pi\Xi}$ and $a_{\bar{K}\Lambda}$ which minimize $\Delta a_{0}$, we show the density plot of $\Delta a_0$ in the $a_{\pi\Xi}$-$a_{\bar{K}\Lambda}$ plane in Fig.~\ref{fig:deltafdensityplot}. We find the optimized subtraction constants at $a_{\pi\Xi}=-2.90$ and $a_{\bar{K}\Lambda}=0.36$, which gives the $K^{-}\Lambda$ scattering length as 
\begin{align}
a_{0,{\rm{th}}}=-0.27-0.40i \ {\rm{fm}},
\end{align}
which reproduces $a_{0,\rm{ex}}$ in the accuracy of $0.01$ fm. This is called Model 2 in the following. In Table~\ref{tab:subtractioinconstants}, we show the subtraction constants of Model 1 and Model 2 in comparison with those of Set 1 in Ref.~\cite{Ramos:2002xh}.

\begin{figure}[tbp]
    \centering
    \includegraphics[width=8cm]{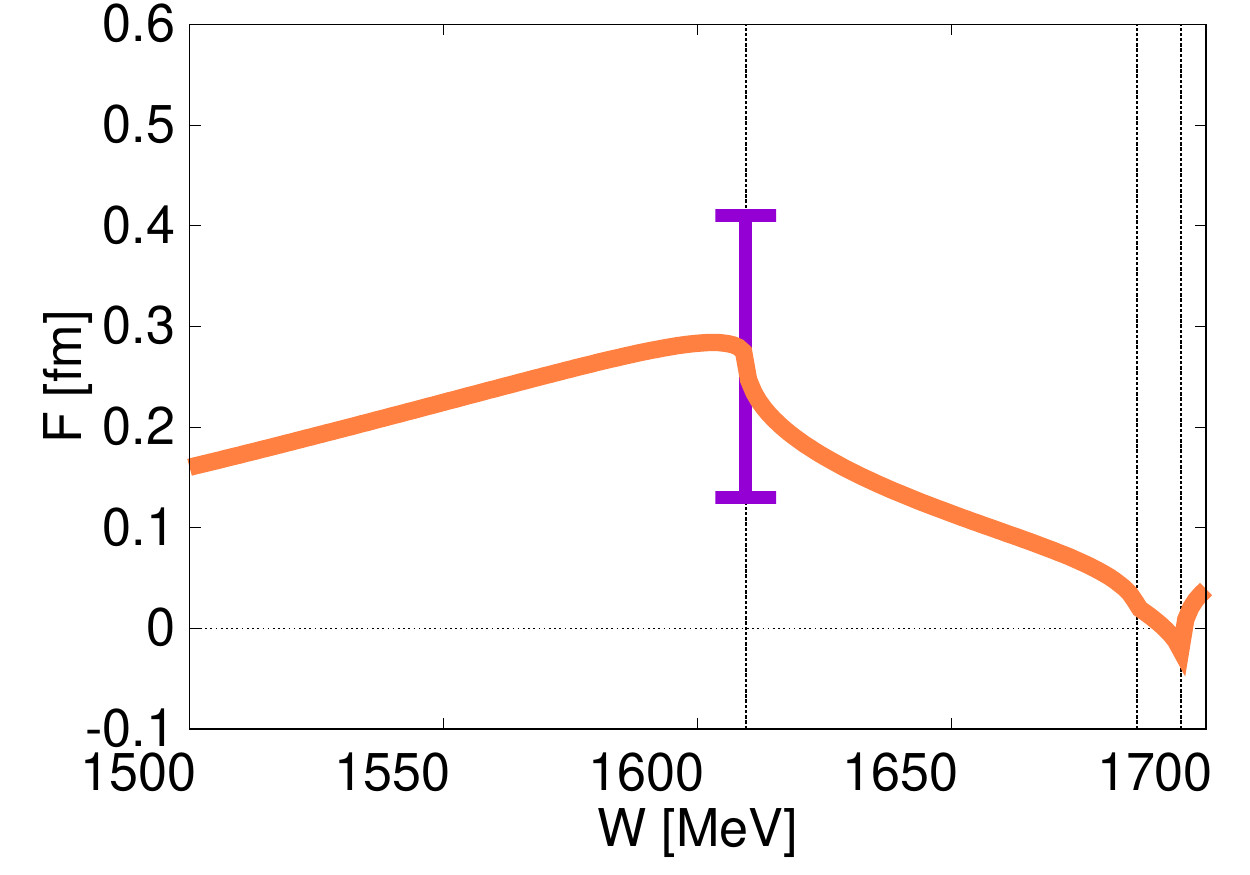}
    \includegraphics[width=8cm]{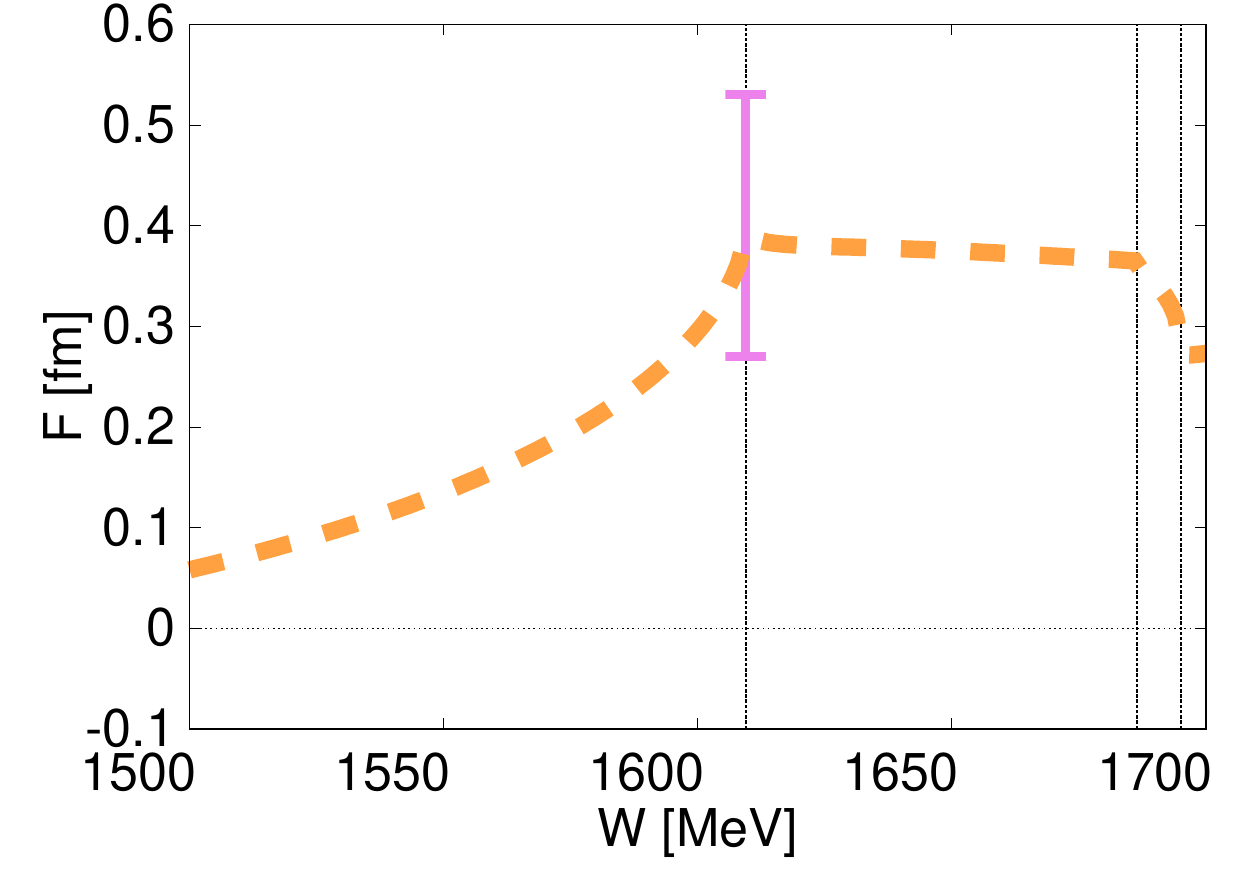}
    \caption{
    $K^{-}\Lambda$ elastic scattering amplitude. 
    Solid (dashed) lines stand for the real (imaginary) parts.
    The real and imaginary parts of the $K^{-}\Lambda$ scattering length by the ALICE collaboration~\cite{ALICE:2020wvi} are shown by the error bars with \eqref{eq:model2space}.
    Vertical dotted lines represent the thresholds of $\bar{K}^0\Lambda$, $K^-\Sigma^+$, $\bar{K}^0\Sigma^0$ from left to right.
    }
  \label{fig:model2realaxis}
\end{figure}

Figure~\ref{fig:deltafdensityplot} and Table~\ref{tab:subtractioinconstants} indicate that the subtraction constants of Model 2 are different from those of Model 1 and Set 1 of Ref.~\cite{Ramos:2002xh}. To study the behavior of the spectrum, we plot the real and imaginary parts of the $K^-\Lambda$ elastic scattering amplitude in Fig.~\ref{fig:model2realaxis} together with the scattering length measured by the ALICE collaboration indicated by the error bar.
The $K^-\Lambda$ elastic scattering amplitude of Model 2 goes across the error bars at the threshold, but shows a different behavior from that of Model 1 in Fig.~\ref{fig:compprevandmodel1}. The spectrum obtained in the experiment corresponds to the imaginary part in Fig.~\ref{fig:model2realaxis}, which shows a cusp at the threshold without a clear peak structure. This may be caused by the pole with a large decay width as in Set 1 of Ref.~\cite{Ramos:2002xh}, but may also be accompanied by the quasivirtual pole as discussed in Sec.~\ref{sec:polescat}. 

We therefore search for the pole of the scattering amplitude in the complex energy plane. First, we investigate the physically relevant sheet; the [bbtttt] sheet below the $K^{-}\Lambda$ threshold and the [bbbttt] sheet above the threshold. In this sheet, no pole is found in the energy region $1500\ \mathrm{MeV}\leq\mathrm{Re }W\leq1700\ \mathrm{MeV}$ and $0\ \mathrm{MeV}\leq|\mathrm{Im }W|\leq100\ \mathrm{MeV}$. Next, we look for the pole in the [bbtttt], [bbbttt], and [ttbttt] sheets in the energy region $1400\leq\mathrm{Re}W\leq1700\ \mathrm{MeV}$ and $0\leq|\mathrm{Im}W|\leq300\ \mathrm{MeV}$. A pole is found in the [ttbttt] at\footnote{In the [ttbttt] sheet, there is also a pole at $1655-53i$ MeV. As discussed in Sec.~\ref{sec:polescat}, the pole with a negative imaginary part corresponds to the $\overline{QV}$ state. In Eq.~\eqref{eq:polea}, we present the pole in the upper half plane which is connected to $R$ and $QB$ continuously.
} 
\begin{align}\label{eq:polea}
z=1655+53i\ \mathrm{MeV},
\end{align}
which corresponds to the quasivirtual ($QB$) state discussed in Sec.~\ref{sec:polescat}.

\subsection{$K^-\Lambda$ scattering length and $\Xi(1620)$ pole} \label{subsec:polelength}

As shown in Sec.~\ref{sec:polescat}, the pole near the threshold is closely related to the scattering length. In this section, we discuss the nature of the $\Xi(1620)$ state in Model 1 and Model 2 in relation with the $K^{-}\Lambda$ scattering length. In Table~\ref{tab:polestat}, we summarize the scattering length $a_{0}$ and its magnitude $|a_{0}|$ of Model 1 and Model 2. It is seen that Model 1 has larger magnitude $|a_{0}|$, reflecting the closer $\Xi(1620)$ pole to the threshold. From the discussion in Sec.~\ref{sec:slength}, when the magnitude of $|a_{0}|$ is sufficiently large, the energy of the pole $z$ can be estimated as 
\begin{align}
z=\frac{-1}{2\mu_{K^-\Lambda}a^2_0}+M_\Lambda+m_{K^-} \label{eq:az}
\end{align}
where $\mu_{K^-\Lambda}=1/[1/m_{K^-}+1/M_{\Lambda}]$ is the reduced mass of the $K^-\Lambda$ system.

The pole positions estimated by Eq.~\eqref{eq:az} from the scattering length are also shown in Table~\ref{tab:polestat} together with the exact pole positions in Eqs.~\eqref{eq:model1poleQm1} and \eqref{eq:polea}. The estimated result of Model 1 ($1615-38i$ MeV) is obtained close to the exact pole position at $1611-30i$ MeV. This reflects the large magnitude of the scattering length. The deviation of about 10 MeV is caused by the contribution from the higher order terms in the effective range expansion. On the other hand, Eq.~\eqref{eq:az} gives a pole on the [ttbttt] sheet, because both the real and imaginary parts of the scattering length of Model 2 are negative. In this case, the discussion in Sec.~\ref{sec:polescat} indicates that  the pole is not on the physically relevant Riemann sheet. This is in agreement with the exact pole in the [ttbttt] sheet found in Sec.~\ref{subsec:Model2}. At the same time, the position of the estimated pole is substantially deviated from the exact one. In order for the estimation~\eqref{eq:az} to work, the magnitude of the scattering length should be sufficiently large. In Model 2, the magnitude is as small as 0.48 fm, which causes the deviation of about 180 MeV.

\begin{table*}[tbp]
\centering
\caption{Scattering lengths $a_{0}$ and the pole positions of Model 1 and Model 2.
}
\label{tab:polestat}
\begin{ruledtabular}
\begin{tabular}{c c c c c}
 & $a_0$ [fm] & $|a_0|$ [fm] & $z_{\rm th}$ [MeV] & 
$z$ estimated by Eq.~\eqref{eq:az} [MeV] \\
\hline
Model 1 & $0.80-0.92i$ & $1.21$ & $1609-30i$  [bbtttt] & $1615-38i$ [bbtttt] \\
Model 2 & $-0.27-0.40i$ & $0.48$ & $1655+53i$  [ttbttt] & $1701+228i$ [ttbttt]
\end{tabular}
\end{ruledtabular}
\end{table*}

\subsection{Comparison of Model 1 with Model 2} \label{subsec:compmodel12}

We have constructed Model 1 which has the $\Xi(1620)$ quasibound state with a narrow width as indicated by the Belle data in Sec.~\ref{subsec:Model1} and 
Model 2 which reproduces the $K^-\Lambda$ scattering length of the ALICE measurement in Sec.~\ref{subsec:Model2}. While both Model 1 and Model 2 are based on experimental data, the fitted values of the subtraction constants show a sizable difference. Here we try to construct a model which satisfies both the constraints by taking into account the experimental uncertainties.

\begin{figure}[tbp]
\centering
\includegraphics[width=8cm]{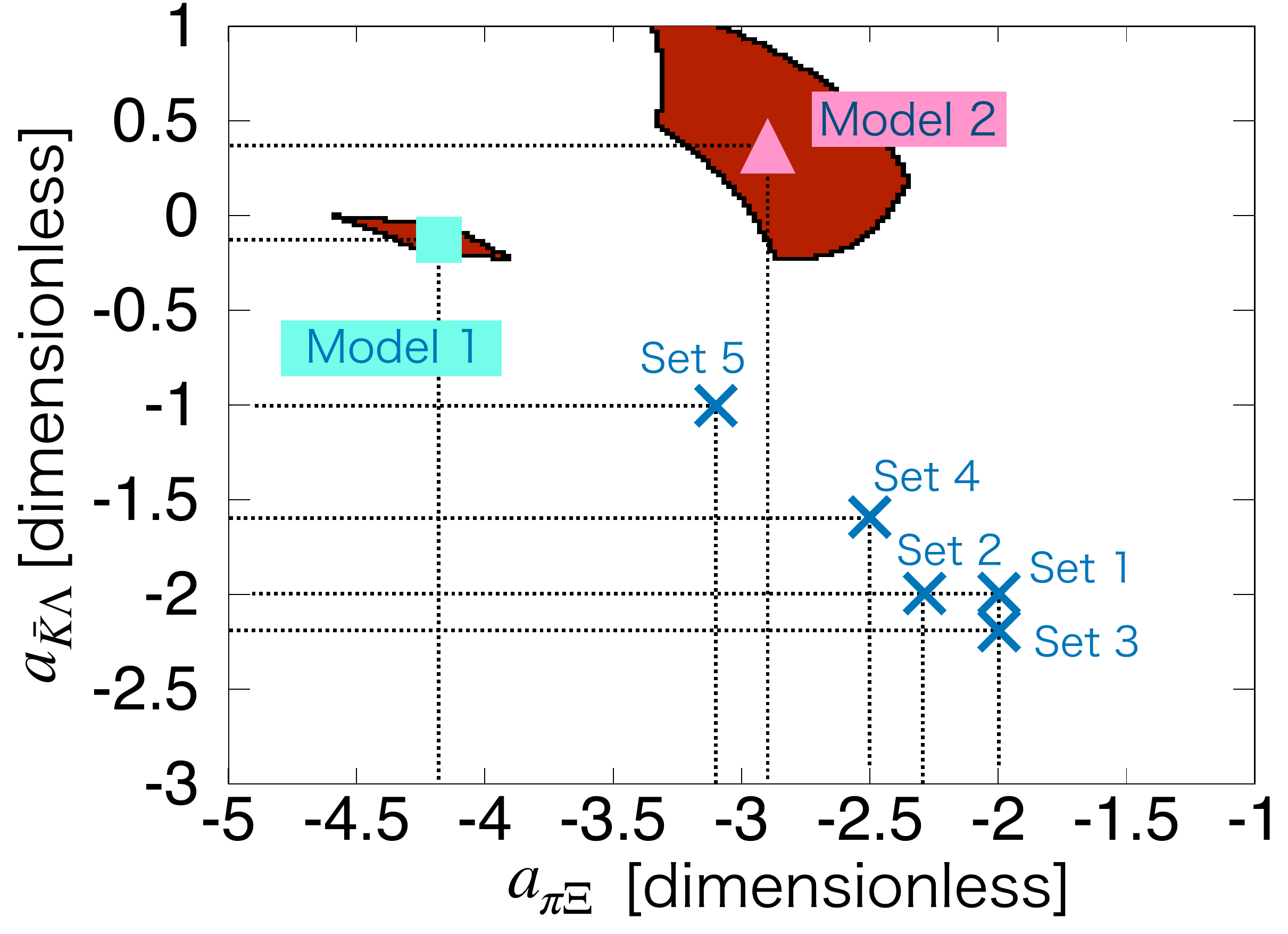}
\caption{
Comparison of the parameter regions of the models with Eq.~\eqref{eq:model1space} and those with Eq.~\eqref{eq:model2space}
in the $a_{\pi\Xi}$-$a_{\bar{K}\Lambda}$ plane.
Triangle, square, and circle denote the parameters of Model 2, Model 1, and models in Ref.~\cite{Ramos:2002xh}.
}
\label{fig:model12}
\end{figure}

First, we determine the constraints on the pole position by taking the quadrature of the statistical and systematic uncertainties of the Belle result~\eqref{eq:Belleresult}:
\begin{align}\label{eq:model1space}
\begin{split}
1603.1\ {\rm{MeV}}&\leq {\rm{Re}}\ z_{\rm{th}}\leq1616.5\ {\rm{MeV}},
\\{\rm{and}} \quad 25.7\ {\rm{MeV}}&\leq {\rm{Im}}\ z_{\rm{th}}\leq32.8\ {\rm{MeV}},
\end{split}
\end{align}
In the same way, the $K^-\Lambda$ scattering length from Eq.~\eqref{eq:resultalice5} is
\begin{align}\label{eq:model2space}
\begin{split}
-0.41\ {\rm{fm}}\leq {\rm{Re}}\ a_{0,\mathrm{th}}\leq-0.13\ {\rm{fm}},
\\{\rm{and}} \quad -0.53\ {\rm{fm}}\leq {\rm{Im}}\ a_{0,\mathrm{th}}\leq-0.27\ {\rm{fm}},
\end{split}
\end{align}
In Fig.~\ref{fig:model12}, we show the region of the subtraction constants which gives the pole satisfying Eq.~\ref{eq:model1space} and the region with the $K^-\Lambda$ scattering length satisfying Eq.~\eqref{eq:model2space}. For comparison, we also plot the subtraction constants of Set 1-5 of Ref.~\cite{Ramos:2002xh}. We find that there is no solution which satisfies both the pole position of Eq.~\eqref{eq:model1space} and the $K^{-}\Lambda$ scattering length in Eq.~\eqref{eq:model2space}. Based on the analysis in Sec.~\ref{sec:polescat} and the results in Sec.~\ref{subsec:polelength}, we expect that this discrepancy is rooted to the nature of the quasibound state ($QB$) pole in Eq.~\eqref{eq:model1space} and the sign of the real and imaginary parts of the scattering length in Eq.~\eqref{eq:model2space}. Thus, we conclude that a narrow width $\Xi(1620)$ quasibound state slightly below the $\bar{K}\Lambda$ threshold is not compatible with the $K^{-}\Lambda$ scattering length measured by the ALICE collaboration. We also note that the Belle result~\eqref{eq:Belleresult} is obtained from the fit to the experimental $\pi\Xi$ spectrum by the Breit-Wigner distribution. For more quantitative comparison, it is necessary to perform the fit to the $\pi\Sigma$ spectrum directly, rather than the assumption of the pole position as in Eq.~\ref{eq:model1space}.

\section{Summary}

In this work, focusing on $\Xi(1620)$ for which the new experimental constraints are being available, we study the nature of the near-threshold state and the scattering length using the numerical calculations with the chiral unitary approach. In Sec.~\ref{sec:polescat}, we discuss the eigenstate represented by the near-threshold pole in the complex energy plane from the viewpoint of the scattering length in a two-channel problem. We show that the resonance above the threshold $R$ and the quasibound state below the threshold $QB$ are not directly connected due to the structure of the Riemann sheets, and in between $R$ and $QB$ there is a pole on the [tb] sheet which can be called the quasivirtual state $QV$. We show that the pole representing $QB$, $QV$, and $R$ can be classified by the sign of the real and imaginary parts of the scattering length. In particular, the resonance $R$ above the threshold cannot be expressed by only the scattering length, and the contribution from the effective range is necessary.

In Sec.~\ref{subsec:Model1}, we construct Model 1 which reproduces the pole of the quasibound state as indicated by the Belle result. It is found that the imaginary part of the $\pi^+\Xi^-$ scattering amplitude shows a peak of $\Xi(1620)$ in Model 1. At the same time, the threshold effect is quantitatively demonstrated through the comparison with the Breit-Wigner amplitude which has a pole at the same position.

Model 2 is constructed in Sec.~\ref{subsec:Model2} by reproducing the $K^-\Lambda$ scattering length by the ALICE experiment. We find that the $K^{-}\Lambda$ scattering amplitude shows a cusp structure at the threshold in Model 2. The pole of $\Xi(1620)$ is not in the physically relevant Riemann sheet, but in the [ttbttt] sheet as a quasivirtual state.

In Sec.~\ref{subsec:polelength}, we estimate the pole positions from the $K^{-}\Lambda$ scattering lengths of Model 1 and Model 2 as an application of the discussion in Sec.~\ref{sec:polescat}. As a result, we find that the estimated pole appears in the same Riemann sheet with the exact pole both for Model 1 and Model 2. Quantitatively, the pole is estimated in a 10 MeV precision for Model 1 where the absolute value of the scattering length is of the order of 1 fm, while Model 2 with a smaller magnitude of the scattering length shows the deviation of about 180 MeV.

In Sec.~\ref{subsec:compmodel12}, we examine the compatibility of the assumption of the quasibound $\Xi(1620)$ state in Model 1 and the $K^-\Lambda$ scattering length by the ALICE measurement in Model 2 including the experimental uncertainties. It is found that the existence of a narrow quasibound state is not compatible with the $K^{-}\Lambda$ scattering length even with the experimental uncertainties. This is because the $K^{-}\Lambda$ scattering length by the ALICE collaboration indicates a quasivirtual pole in the [ttbttt] sheet, while the quasibound state in the [bbtttt] sheet indicates the positive real part of the scattering length $a_{0}$. At the same time, it should be kept in mind that the Belle data does not directly constrain the pole position of the scattering amplitude, and the near-threshold pole should be determined together with the threshold effect as shown in this paper.

As future perspectives, the calculation of the $\pi^+\Xi^-$ invariant mass spectrum in the $\Xi_c\rightarrow\pi^{+}\pi^{+}\Xi^{-}$ decay, as was done in Ref.~\cite{Miyahara:2016yyh}, is important for the comparison with the Belle data. This may provide a description of the Belle data in a consistent way with the ALICE measurement of the $K^{-}\Lambda$ scattering length. In this paper, we have discussed the $\bar{K}\Lambda$ threshold effect on the $\Xi(1620)$ state. Similar threshold effect is expected for $\Xi(1690)$ whose peak is located near the $\bar{K}\Sigma$ threshold at $1686.1\ {\rm MeV}$. In this case, the threshold effect should be much more complicated because there are two thresholds ($K^-\Sigma^+$ and $\bar{K}^0\Sigma^0$ for the neutral channel) due to the isospin symmetry breaking. It is also important to investigate the internal structure of $\Xi(1620)$ based on the models constructed in this paper. The internal structure can be extracted, for instance, by the method of the natural renormalization scheme in Ref.~\cite{Hyodo:2008xr} and by the evaluation of the compositeness through the residue of the resonance pole~\cite{Hyodo:2011qc,Sekihara:2014kya}.

\begin{acknowledgments}
The authors thank Wren Yamada for the useful discussion on the Riemann sheets and 
the pole of the coupled-channel scattering.
This work has been supported in part by the Grants-in-Aid for Scientific Research from JSPS (Grant numbers
JP22K03637, % Kiban C (Hyodo)
JP19H05150, % Cluster (Hyodo)
JP18H05402). % Cluster buntan (Hyodo)
This work was supported by JST, the establishment of university
fellowships towards the creation of science technology innovation, Grant Number JPMJFS2139. 
\end{acknowledgments}

\bibliography{refs}

\end{document}